\documentclass[prb,superscriptaddress,twocolumn,showkeys]{revtex4-2}
\usepackage{graphicx}
\usepackage{hyperref}
\usepackage{amsmath}
\usepackage{amssymb}
\usepackage{bm}
\usepackage{units}
\usepackage{xcolor}
\newcommand{\rem}[1]{}

\newcommand{\boldepsilon}{\boldsymbol{\epsilon}}

\DeclareMathOperator{\RE}{\rm Re}
\DeclareMathOperator{\IM}{{\rm Im}}

\begin{document}

\title{Phononic frictional losses of a particle crossing a crystal: linear-response theory}

\author{Gabriele Riva}
\affiliation{Dipartimento di Fisica, Universit\`a degli Studi di Milano,
  Via Celoria 16, 20133 Milano, Italy}

\author{Giacomo Piscia}
\affiliation{Dipartimento di Fisica, Universit\`a degli Studi di Milano,
  Via Celoria 16, 20133 Milano, Italy}

\author{Nicolas Trojani}
\affiliation{Dipartimento di Fisica, Universit\`a degli Studi di Milano,
  Via Celoria 16, 20133 Milano, Italy}

\author{Giuseppe E. Santoro}
\affiliation{International School for Advanced Studies, Via Bonomea 265,
  34136 Trieste, Italy}  
\affiliation{The Abdus Salam International Center for Theoretical Physics, Strada Costiera 11, 34151 Trieste, Italy}
\affiliation{CNR-IOM Democritos National Simulation Center,
  Via Bonomea 265, 34136 Trieste, Italy}

\author{Erio Tosatti}
\affiliation{International School for Advanced Studies, Via Bonomea 265,
  34136 Trieste, Italy}
\affiliation{The Abdus Salam International Center for Theoretical Physics, Strada Costiera 11, 34151 Trieste, Italy}
\affiliation{CNR-IOM Democritos National Simulation Center,
  Via Bonomea 265, 34136 Trieste, Italy}

\author{Nicola Manini}
\affiliation{Dipartimento di Fisica, Universit\`a degli Studi di Milano,
  Via Celoria 16, 20133 Milano, Italy}

\begin{abstract}
   We address weak-coupling frictional sliding with phononic dissipation
   by means of analytic many-body techniques.
   Our model consists of a particle (the "slider")  moving through a two- or three-dimensional crystal and interacting weakly with its atoms, and therefore exciting
   phonons.
   By means of linear-response theory we
   obtain explicit expressions for the friction force slowing down the slider as a function of its speed, and compare them to the friction obtained by simulations, demonstrating a remarkable accord.
\end{abstract}

\keywords{Friction, sliding friction, kinetic friction, linear-response theory}

\maketitle

%----------------------------------------------------------------------------
\section{Introduction}\label{intro}
%----------------------------------------------------------------------------
Nanofriction investigates the complex processes of transformation of
mechanical energy into thermal energy, 
through excitation of the
vibrational degrees of freedom (phonons) of solids, as well as electronic
ones, when available.
Friction phenomena between dry bodies occur on multiple length and time
scales, from the macroscopic scale to the atomic scale.
Also, they span from weak to strong magnitudes, from gentle wearless sliding to heavy, wear-rich scratching/ploughing.
The present paper addresses the quantitative theory of a special
wear-free friction at the atomic scale,
namely that of weak-coupling phononic dissipation for a point-like classical particle channeling through a bulk crystal.
The current experimental interest in the investigation of the detailed
speed dependence of phononic friction \cite{Tao23,Huang24} indicates that a
general microscopic theory of the phonon coupling and dissipation is called
for. 

As a preamble, it must be noted that dynamical friction is rarely accessible to linear-response (LR) theory, because the coupling between "slider" and "substrate" and its effects are mostly far from weak, and the response elicited far from linear.

However, a weak coupling situation in which LR works is realized in {\it two} well-defined limits. 
One is the popular low-velocity, high temperature limit of a nanoscale slider. 
In that limit the thermal random walk of an initially freely diffusing microscopic slider is weakly perturbed by an infinitesimal driving force field.
The forcing gives rise to a mean average drift velocity with a
corresponding frictional dissipation, both infinitesimal, connected to the
initial random walk parameters through Einstein's fluctuation-dissipation
theorem \cite{Einstein05,Callen51,Kubo66}.
This frictional regime was addressed, e.g., in Ref.~\cite{Smith96}.
The second limit, opposite to the first, is realized when the slider's own velocity, momentum, or mass are so large that the disturbance caused to its motion can be altogether neglected. 
Such is, for instance, the general framework assumed in the dielectric
theory of energy losses by fast particles crossing matter, long established
for electronic excitations \cite{LandauEMv8_ch14,Pines99}.
In the frictional context, the two opposite limits of low and large slider
velocity were demonstrated in a molecular-dynamics (MD) simulation of
%a similar system,
a gold cluster sliding on a graphene substrate, where the
fast-slider limit was designated as ``ballistic'' \cite{Guerra10}.

The present theory addresses the same fast-slider limit, focusing on phonon
losses (rather than electronic ones).
It extends to a fast particle crossing, nearly unscathed, a 2D or 3D
crystal, the results obtained earlier for a simple 1-dimensional (1D) model
\cite{Panizon18,Panizon24erratum}.
Friction is evaluated as the loss of energy per unit time of a
channeling particle that interacts weakly, through conservative short-range
forces, with a vibrating crystal.
The sliding particle, which in the following we still refer to as the ``slider'', traverses the crystal,
generates phonon
excitations, which we describe at a quantum level. 
The end result is the following formula for the friction force as a
function of the slider velocity $v_\text{SL}$ and the slider--crystal-atom
interaction potential $V(r)$, or rather its Fourier transform
$\tilde{V}({\mathbf Q})$:
\begin{widetext}
\begin{align}\label{friction_dissfinal:eq}
  F(v_\text{SL})=&
  \frac{4\pi}{m a^3} \sum_{\mathbf G_\perp} 
  e^{-i \mathbf{x}_0 \cdot \mathbf G_\perp} 
  \int_\Omega \frac{d^3Q}{(2\pi)^3} \, Q_x \,
%  \\\nonumber  &\times
  \tilde{V}(\left| \mathbf Q \right|)\,
  \tilde{V} (\left| \mathbf Q + \mathbf G_\perp
  \right|) \, e^{-W(\mathbf{Q})-W(\mathbf{Q}+\mathbf{G}_\perp)} \times
 \\ \nonumber  
 & \sum_{\lambda}
  \mathbf{Q} \cdot\boldepsilon_{\lambda}(\mathbf{Q}) \,
    (\mathbf{Q} +\mathbf{G}_\perp)\cdot
    \boldepsilon_{\lambda}(\mathbf{Q})\,
%  \mathcal{L}(\mathbf{Q}, v_\text{SL}, \gamma) \,
  \frac{\gamma}{2\pi}\,
  \frac{4 Q_x v_\text{SL}}
  {[(Q_x v_\text{SL} -
    \omega_{\lambda}(\mathbf{Q}))^2 + (\frac{\gamma}{2})^2] \, [(Q_x v_\text{SL} +
    \omega_{\lambda}(\mathbf{Q}))^2 + (\frac{\gamma}{2})^2]}  
  \,.
\end{align}
\end{widetext}
Here the integration over the wave vector $\mathbf{Q}$ extends over the octant $\Omega$ with all positive components. $\omega_{\lambda}(\mathbf{Q})$ and $\boldepsilon_{\lambda}(\mathbf{Q})$ are the crystal's phonon frequencies and polarization vectors, respectively.
$\mathbf{G}_\perp$ are the reciprocal-lattice vectors perpendicular to the direction $\hat{\mathbf{x}}$ of the slider velocity;
$\mathbf{x}_0$ is the slider's initial position;
$W(\mathbf{Q})$ are Debye-Waller factors;
and $\gamma$ is a coefficient quantifying the (weak) dissipation leading to phonon decay in the crystal.
The predictions of Eq.~\eqref{friction_dissfinal:eq} are validated by
striking quantitative agreement with direct energy-loss MD simulations.

This paper is organized as follows.
We introduce the model in Sect.~\ref{model:sec}, detailing in particular the potential-energy function that describes the slider-crystal interaction.
In Sect.~\ref{LRTfriction:sec} we report the theory leading to a general friction formula, Eq.~\eqref{friction_imm:eq}, an 
expression relying on the density-density LR function $\chi_{nn}^R$ of the crystal.
Section~\ref{Force:sec} reports the derivation of Eq.~\eqref{friction_dissfinal:eq},
namely an explicit version of Eq.~\eqref{friction_imm:eq} based on the
evaluation of $\chi_{nn}^R$ using a standard one-phonon approximation and
on the assumption that the crystal comes with an intrinsic dissipation
mechanism.
In Sect.~\ref{3D:sec} we compare the analytic expression Eq.~\eqref{friction_dissfinal:eq} with the friction evaluated numerically through molecular dynamics (MD) simulations.
Section~\ref{2D:sec} reports a version of the friction formula specialized to the 2D square-lattice crystal, also compared with 2D MD simulations.
In Sect.~\ref{discuss:sect} the present results and future
extensions are discussed in relation with the current literature of the field.

\section{The model}\label{model:sec}

\begin{figure} 
\centerline{
\hfill 
\includegraphics[width=0.9\linewidth]{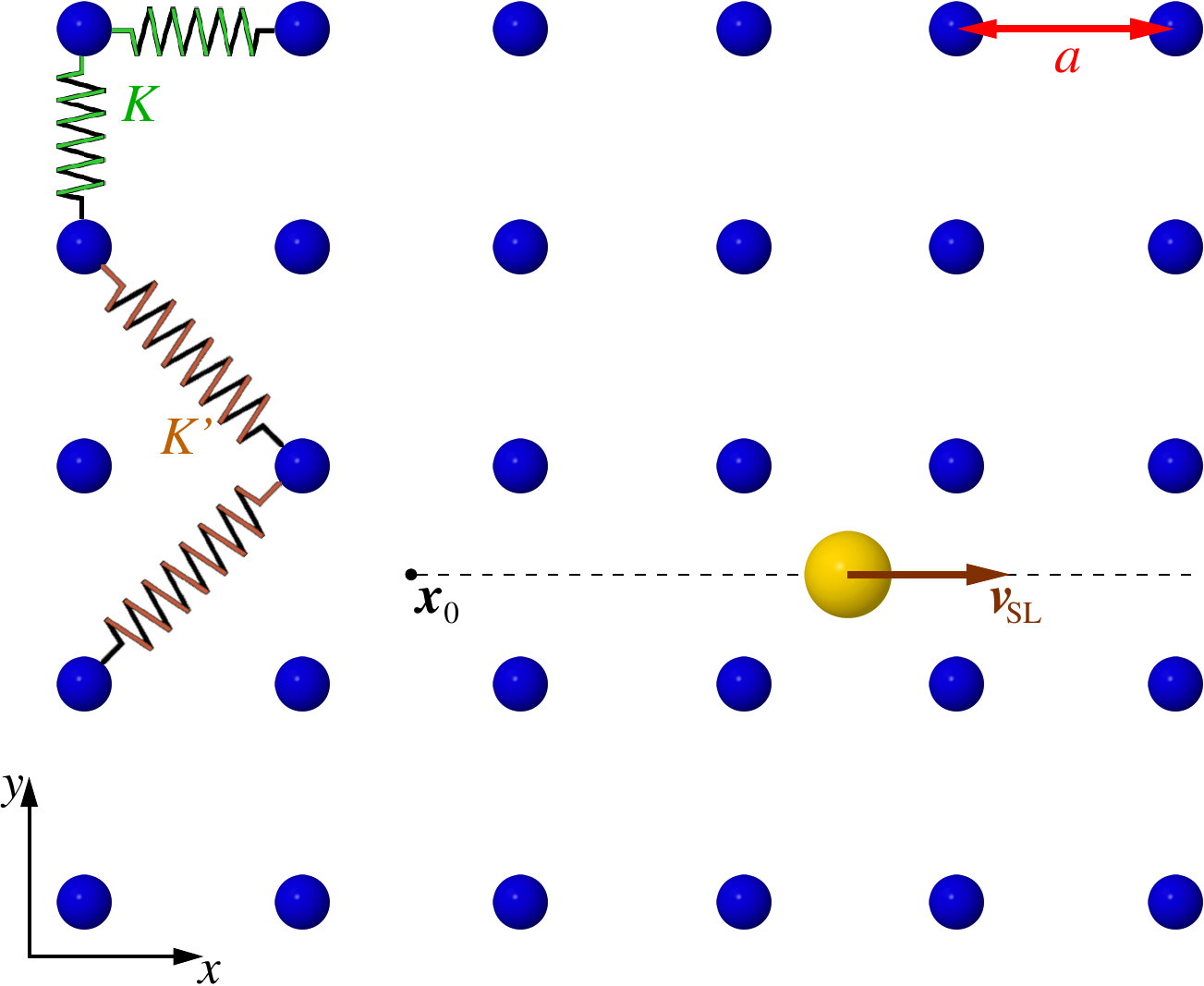}
\hfill
}
\caption{\label{Model:fig}
  A sketch of the model: the slider (large yellow sphere) follows a channeled trajectory (dashed line) inside a crystal.
  The slider interacts with each atom in the harmonic crystal through a 2-body potential $V(r)$.
  The vibrational properties of the simple-cubic (3D) or square (2D) lattice crystal are determined by first- and second-neighbor springs with elastic constants $K$ and $K'$, respectively, see Appendix~\ref{lattice:app}.
}
\end{figure}

A sliding object, such as an atomic-force-microscope (AFM) tip, gently grazing a flat crystal surface is slowed down by frictional phenomena.
The kinetic energy of the slider decreases due to a sequence of
collisions with the crystal, which generate excitations.
In the present work, we omit the direct excitation of electrons and all
kinds of triboelectric phenomena that can occur in real life.
Focusing on insulators, within the adiabatic scheme, we address the
excitation of 
phonons.
These phonons, assumed to be harmonic, will  ballistically propagate away, their energy and momentum lost by the slider.
The realistic condition of a slider in contact with a surface involves
necessarily all complications associated with the breaking of the lattice
discrete invariance, resulting in surface phonons, e.g., Rayleigh waves.
In the present paper we temporarily disregard surface effects, and consider
the idealization of an infinitely extended perfect crystal, traversed by a
slider that ``channels'' through it, as sketched in Fig.~\ref{Model:fig}.

Our model harmonic crystal is characterized by particles of mass $m$ that
at equilibrium are arranged as a square (2D) or simple-cubic (3D) lattice.
Harmonic nearest- and second-neighbor springs with elastic constants $K$
and $K'$ and equilibrium lengths $a$ and $\sqrt{2} a$, respectively,
guarantee the mechanical stability and determine the phonon dispersions.

In its simplest form, the slider is implemented as a point particle,
characterized by mass $M$, position $\mathbf{x}_\text{SL}$ and velocity
$\mathbf{v}_\text{SL}$.
More articulate structures of the slider are certainly possible, and were
tested in the 1D context \cite{Panizon18}.
The model can be formulated as follows:
\begin{align}
  H = H_\text{harm} +
  H_\text{SL} +
  \sum_j V (|\mathbf{x}_j -\mathbf{x}_\text{SL}(t)  |)
  \,,
\end{align}
where $H_\text{harm}$ is the Hamiltonian for the harmonic lattice,
$H_\text{SL} = {\mathbf p}^2_\text{SL}/(2M)$ is the Hamiltonian for
the freely moving point slider, and the interaction couples the slider
motion to all atoms in the harmonic crystal through a 2-body potential
energy function $V(r)$.
This model makes perfect sense for arbitrary kind and strength of this
interaction potential, and for arbitrary channeling direction and velocity
$\mathbf{v}_\text{SL}$.
For general conditions, and whenever
quantum-mechanical effects are
negligible
(such as could be approximately met in a real crystal at moderate temperature)
it is relatively straightforward to investigate, e.g.\ by means of
classical MD simulations involving a (sufficiently large) crystal portion. 
This model describes in microscopic detail the slider-crystal two-way
energy and momentum transfer that, in time, leads to an overall slowing
down of the slider and a progressive uptake of energy (heat) by the
crystal.
The instantaneous kinetic friction force experienced by the slider is quantified by its deceleration multiplied by the slider mass $M$.
The same quantity averaged over a sufficiently long time allows us to
estimate the mean kinetic friction
\begin{align}
  \mathbf{F}(t_1,t_2) =
  -M \frac{\mathbf{v}_\text{SL}(t_2)-\mathbf{v}_\text{SL}(t_1)}{t_2-t_1}
\end{align}
that slows down the slider motion.

Aiming at an analytic evaluation of this mean kinetic friction, we
introduce a few simplifying assumptions/approximations.
The main assumption is that the interaction between the slider and the
crystal is so weak that it perturbs the slider motion only over very long
time scales.
This allows us to adopt the first Born approximation of scattering
theory: the slider moves at constant velocity
\begin{equation}\label{xSL:eq}
\mathbf{x}_\text{SL}(t) = \mathbf{x}_0 + \mathbf{v}_\text{SL} t
\,,
\end{equation}
as was done in the investigation of a similar 1D model \cite{Panizon18}.
For the validity of this assumption, the kinetic energy of the particle,
$\frac 12 Mv_\text{SL}^2$, must be much larger than the typical energy
transferred in the time taken by the slider to advance by one lattice spacing.
This observation implies that, to be quantitatively valid, the adopted
first Born approximation requires not only that the interaction strength is
small, but also that the slider speed $v_\text{SL}$ and mass $M$ are
sufficiently large.

We also assume that the slider moves along a crystal high-symmetry
direction, say the $x$ axis: $\mathbf{v}_\text{SL} =
v_\text{SL}\mathbf{e}_x$.
We also postulate that it starts off at a symmetric position $\mathbf{x}_0$
inside the crystalline ``channel'', e.g.\ exactly midway between lines of atoms.
In 2D, we take the symmetric slider's initial position $\mathbf{x}_0=\frac
a2\,\mathbf{e}_y$, i.e.\ equidistant from the two adjacent atomic rows in
the $\mathbf{e}_x$ direction of the slider motion, see Fig.~\ref{Model:fig}.
In the 3D simple-cubic lattice, for $\mathbf{x}_0$  we test the following high-symmetry starting points in the $y-z$ plane: either at the centre of a square,
$\mathbf{x}_0=\frac a2(\mathbf e_y+\mathbf{e}_z)$, or at the midpoint of a bond, $\mathbf{x}_0=\frac a2\, \mathbf{e}_y$.
As we target a steady state, our analysis is independent of the component
of the initial position in the sliding direction $\mathbf{e}_x$.

\begin{table}
    \begin{center}
    \begin {tabular}{c|c|c}
    \hline
    \hline
    Physical quantity & natural units & typical values \\
    \hline
    length & $a$ & $500$~pm\\
    mass & $m$ & $5 \times 10^{-26}$~kg\\
    spring constant & $K$ & $300 $~N/m\\
    time & $(m/K)^{1/2}$ & $1.3 \times 10^{-14}$~s \\
    frequency & $(K/m)^{1/2}$ & $7.7 \times 10^{13}$~s \\
    velocity & $a(K/m)^{1/2} $ & $3.9 \times 10^4$~m/s\\
    force    & $Ka$ & $1.5 \times 10^{-7}$~N\\
    energy & $Ka^2$ & $7.5 \times 10^{-17}$~J\\
    temperature & $Ka^2/k_\text{B}$ & $5.4 \times 10^{6}$~K\\
    action & $a^2 (Km)^{1/2}$ & $9.7 \times 10^{-31}$~Js\\
    \hline
    \hline
    \end{tabular}
    \end{center}
    \caption{\label{uofm:tab}
    Natural model units for the relevant physical quantities, with
    indications of plausible values relevant for a standard crystal.
    The natural energy scale represents the typical classical potential energy required to displace one atom away from its equilibrium position by one entire lattice spacing $a$, a quantity comparable to the crystal cohesive energy, and much larger than typical vibrational phonon quanta $\hbar\omega$.
    }
\end{table}

As was done for the 1D problem \cite{Panizon18}, it is convenient to
express all mechanical quantities in natural units related to the crystal
dynamics, listed in Table~\ref{uofm:tab}.
The main difference with the 1D model, where the unique speed of sound provides the natural unit of velocity, is that the 2D/3D crystals have multiple (transverse and longitudinal) speeds of sound, evaluated in Appendix~\ref{lattice:app}.
Nonetheless, all these speeds of sound are of the order of $a(K/m)^{1/2} $, which we adopt as the typical velocity unit, as indicated in Table~\ref{uofm:tab}.

Within the discussed approximations, the slider dynamics becomes trivial,
thus irrelevant.
Accordingly, the Hamiltonian can be rewritten as that of a harmonic crystal weakly perturbed by the time-dependent interaction generated by the slider:
\begin{align}
  H &= H_\text{harm} + \sum_j
  V (|\mathbf{x}_j - \mathbf{x}_0 -\mathbf{v}_\text{SL} t  |)
    \\\nonumber
    &= H_\text{harm} + \int d^3x \;
    V_\text{ext}(\mathbf{x},t)\; n(\mathbf{x})
    \,.
\end{align}
Here we have introduced the density operator
\begin{align} \label{densop:eq}
  n(\mathbf{x})=\sum_j \delta^{(3)}(\mathbf{x} - \mathbf{x}_j)
  \,,
\end{align}
involving the position operator $\mathbf{x}_j$  of the $j$-th atom,
and the (weak) interaction  potential energy $V_\text{ext}(\mathbf{x},t)\equiv V(|
\mathbf{x} - \mathbf{x}_0 -\mathbf{v}_\text{SL} t |)$
at a generic
location $\mathbf{x}$ in the crystal, as generated
by the slider instantaneously visiting the
location $\mathbf{x}_0 + \mathbf{v}_\text{SL} t$ at time $t$.
This formulation is fully suitable for the application of the perturbative methods of quantum mechanics.

\subsection{The interaction potential}\label{pot:sec}

The potential mediates the energy transfer to the harmonic crystal.
It needs to be weak to make the perturbative approach meaningful.
Note that, contrary to classical mechanics, in a quantum crystal the atomic
positions are delocalized in space.
The quantum probability distribution for $\mathbf{x}_j$ spreads out across all space.
As a consequence, the potential function $V(r)$ is evaluated at arbitrary
values of its argument $r=| \mathbf{x}- \mathbf{x}_\text{SL}(t)|$.
In particular, this separation $r$ can take values arbitrarily close to
zero, even if the slider remains ``channeled'' midway
between atomic rows.
For this reason, a weak-coupling theory is incompatible with an interaction
that diverges rapidly for $r \to 0$, such as e.g.\ a 6--12 Lennard-Jones
(LJ) potential.
Such kind of rapid divergence would yield a diverging rate of collision
between the slider and a crystal atom, regardless of how small the energy
prefactor in $V$ may be.
Accordingly, our assumptions of weak interaction and constant velocity
would be inappropriate for a sharply diverging potential function.
A short-distance divergence $V\propto r^{-1}$ would be perfectly acceptable instead.
Incidentally, note that the short-distance sharp divergence of the 6--12
LJ function is a realistic model for real-life atom-atom interactions down
to small distances of the order of a fraction of the 1s shell radii, but
below this distance any atom-atom interaction would transition to a
milder $\propto r^{-1}$ divergence.

\begin{figure} 
\centerline{
\includegraphics[width=\linewidth,clip=]{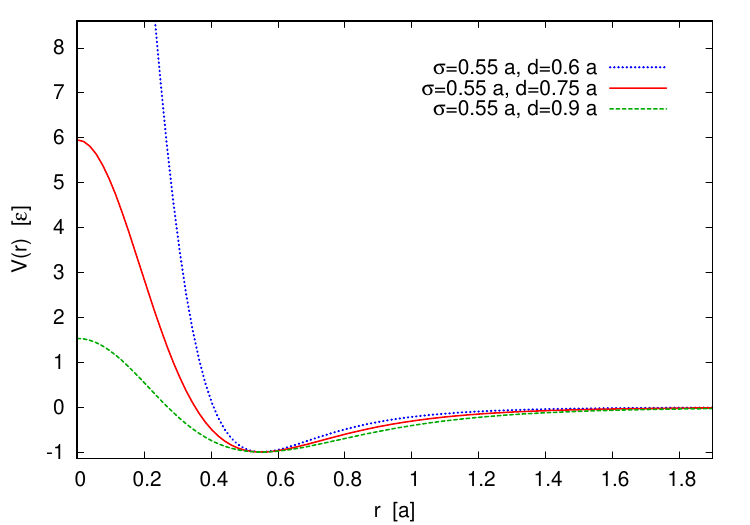}
}
\caption{\label{potx:fig}
  The modified LJ function, Eq.~\eqref{potr:eq}, for $\sigma=0.55 a$ and a few values of the regularization length $d$.
  In the $d\to0$ limit, the standard LJ potential is retrieved.
}
\end{figure}

For the purpose of the present work, where we do not address any specific
real-life condition, we decide to adopt a potential that remains finite
even at close distance.
In practice we adopt a regularized LJ function, defined by:
\begin{equation}\label{potr:eq}
V(r)= \varepsilon
\left[\left(\frac{\sigma^2+d^2}{r^2+d^2} \right)^6
    -2\left(\frac{\sigma^2+d^2}{r^2+d^2} \right)^3
    \right]
.
\end{equation}
Figure~\ref{potx:fig} displays $V(r)$ for a few values of its parameters $\sigma$ and $d$.
This function has a minimum at distance $r=\sigma$ with depth $-\varepsilon$.
$d$ is a regularizing length, a parameter which makes $V(r)$
finite at any $r$.
We adopt values of $d$ of the same order as $\sigma$, so that the potential
does not exceed a few times $\varepsilon$, even at small distance.
The function in Eq.~\eqref{potr:eq} is just a convenient example of model
interaction potential: any sufficiently regular function that vanishes fast
enough at large distance could be adopted in its place, e.g.\ a Yukawa potential, a Gaussian, a Woods-Saxon profile, etc.

\begin{figure} 
\centerline{
\includegraphics[width=\linewidth,clip=]{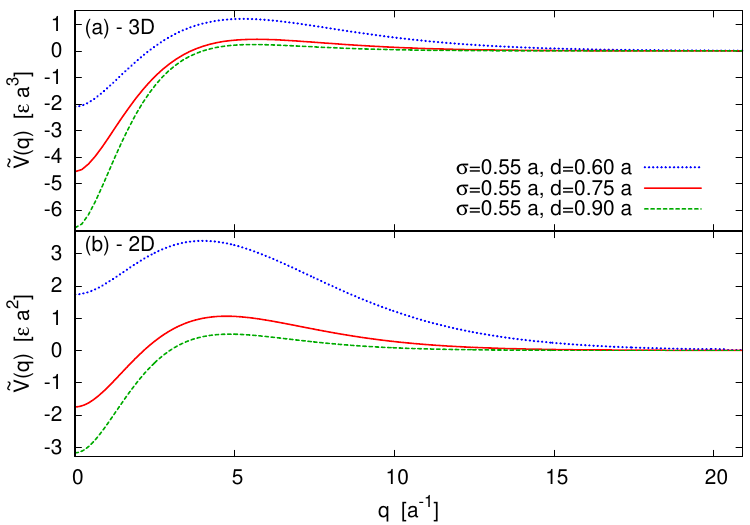}
}
\caption{\label{potq:fig}
The Fourier-transformed regularized
LJ potential of Eq.~\eqref{potr:eq}, for the same values of
$\sigma$ and $d$ as in Fig.~\ref{potx:fig}.
(a): 3D FT, Eq.~\eqref{pot_fourier:eq};
(b): 2D FT, Eq.~\eqref{2D_pot:eq}.
These functions enter the final expressions for friction, e.g.\ Eq.~\eqref{friction_dissfinal:eq} as crucial ingredients.
}
\end{figure}

According to Eq.~\eqref{friction_dissfinal:eq}, and in analogy to the 1D
result of Ref.~\cite{Panizon18}, this interaction $V(r)$ affects
friction through its Fourier transform (FT), which needs to be carried out
taking the dimensionality of the problem into account.
Appendix~\ref{FT3d:appendix} reports the calculation of the 3D FT of the function in Eq.~\eqref{potr:eq}:
\begin{align}\label{pot_fourier:eq}
    \tilde{V}(q)=&\,
    \varepsilon\, \frac{ \pi^2 (\sigma^2+d^2)^3}{2d^3} \, e^{-qd} \,
    \biggl[ \frac{(\sigma^2+d^2)^3}{960d^6}
      \,(q^4d^4
      \\\nonumber &
      +10q^3d^3+45q^2d^2+105qd+105 )
%    \\\nonumber &
      -(qd+1) \biggl]
    \,.
\end{align}
Figure~\ref{potq:fig}a displays this FT for the same values of the parameters $\sigma$ and $d$ adopted for the real-space curves shown in Fig.~\ref{potx:fig}.

Appendix~\ref{FT2d:appendix} reports the derivation of the 2D FT, leading to the following expression:
\begin{align}\nonumber
    \tilde{V}(q)
    =
    2\pi \varepsilon d^2 
    \Big[ &
    \frac 1{5!2^5}
    \left(1+\frac {\sigma^2}{d^2}\right)^6
    (qd)^5 K_5(qd)
    \\\label{2D_pot:eq}&
    -\frac 14
    \left(1+\frac {\sigma^2}{d^2}\right)^3
    (qd)^2 K_2(qd)
    \Big]
    \,,
\end{align}
involving modified Bessel functions of the second kind $K_n(\cdot)$.
Figure~\ref{potq:fig}b illustrates the 2D FT of
Eq.~\eqref{2D_pot:eq}, for the same parameter values as in Fig.~\ref{potx:fig}.

\section{Dissipation and friction}
\label{LRTfriction:sec}

We extend the LR approach carried out for the 1D model \cite{Panizon18,
  Panizon24erratum}, to derive analytic expressions, the main example of
which being Eq.~\eqref{friction_dissfinal:eq}, for the friction force
experienced by the slider gently caressing a crystal.
The weak-interaction regime is guaranteed by adopting a small interaction energy prefactor $\varepsilon$ in Eq.~\eqref{potr:eq}.

The average friction force $F$ can be evaluated by means of the dissipated
power
\begin{equation}\label{WFtimesV}
  \bar W =F \, v_\text{SL}
\end{equation}
in the steady regime.
The time average of the dissipated power must be executed over a period $\tau = a/v_\text{SL}$, namely over the time that it takes for the slider
to advance across one lattice cell.
To make this averaging possible, we assume that the slider advances along a
crystal-commensurate direction, specifically $(100)$:
${\mathbf v}_\text{SL} = v_\text{SL} \hat{\mathbf x}$.
This assumption is fairly natural, since in practice channeling usually
tends to self-align along crystal-commensurate directions \cite{Gemmell74}.
If instead the slider advanced in an arbitrary oblique
direction (not identified by a set of integer Miller indexes), then the sliding process would not be periodic in time, and this would complicate the calculation of this time average $\bar W$.

We start from the instantaneous power transferred by the slider to the
crystal due to the excitation of lattice vibrations.
The straightforward generalization of the LR-theory expression Eq.~(3) of
Ref.~\cite{Panizon18} to a 3D geometry is
\begin{align}\label{Wr3D}
  W=\frac{d}{dt}E(t)  \simeq & 
  - \int d^3x  \int d^3x'
  \int_{-\infty}^{\infty} dt' 
\\\nonumber &
V_\text{ext}(\mathbf{x},t) \,\frac{\partial
 \chi_{nn}^R(\mathbf{x},\mathbf{x}',t-t')}{\partial t}
\,V_\text{ext}(\mathbf{x}',t') \,.
\end{align}
$E(t)$ is the internal energy of the perturbed crystal at time $t$.
Here
$V_\text{ext}(\mathbf{x},t)$ is the perturbation produced at time $t$ by the slider following Eq.~\eqref{xSL:eq}.
As a result
$V_\text{ext}(\mathbf{x},t)=V(|\mathbf{x}-\mathbf{x}_0-\mathbf{v}_\text{SL} t|)$.
Also,
$\chi_{nn}^R({\mathbf x},{\mathbf x}';t-t')=-\frac{i}{\hbar}\theta(t-t')      
\langle [\hat{n}({\mathbf x},t),\hat{n}({\mathbf x}',t')] \rangle$   
is the retarded density-density response function of the
unperturbed harmonic crystal \cite{Vignale05}.
In 2D, the expression remains formally the same, except that the $\mathbf{x}$ and
$\mathbf{x}'$ integrations are carried out over the appropriate 2D space.

We conveniently adopt a Fourier representation for the
external potential:
\begin{equation}
  V_\text{ext}(\mathbf{x},t)=
  \int \frac{d^3q}{(2\pi)^3}\, e^{i\mathbf{q}\cdot (\mathbf{x}-\mathbf{x}_0
    - \mathbf{v}_\text{SL} t)} \, \tilde{V}(|\mathbf{q}|)
  \,,
\end{equation}
where the integration becomes $d^2q/(2\pi)^2$ in 2D.
%
%The Fourier transform $\tilde{V}(|\mathbf{q}|)$ was derived in Sec.~\ref{pot:sec}.
%
Taking advantage of the lattice  translation invariance,
we write the Fourier transform of the retarded response function as:
\begin{align} \nonumber
  \chi_{nn}^R (\mathbf{x},\mathbf{x}',t-t') =& \sum_{\mathbf{G}} \int
  \frac{d^3 Q}{(2 \pi)^3}\int_{-\infty}^{+\infty} \frac{d\omega}{2 \pi} e^{-i \omega (t-t')}\times
\\ \nonumber &
e^{i \mathbf{Q}\cdot \mathbf{x}}
\, \chi_{nn}^R(\mathbf{Q},\mathbf{Q+G},\omega)
\, e^{-i (\mathbf{Q+G})\cdot \mathbf{x}'}
\, .
\end{align}
Here the $\mathbf{G}$ vectors are the reciprocal lattice vector of the
crystal.
The sum over $\mathbf{G}= 2 \pi a^{-1}\, (l_x,l_y,l_z)$ is understood as a
sum over the (integer) Miller indexes $l_x$, $l_y$ and $l_z$.

We now average the instantaneous power over one period, and reformulate
Eq.~\eqref{Wr3D} in the $(\mathbf{Q},\omega)$ Fourier domain:
\begin{widetext}
\begin{align}
  \bar{W} &
  = \frac{1}{\tau} \int_0^{\tau} dt \, W(t)
  %\\ \nonumber &
  = - \frac{1}{\tau}  \int_0^{\tau} dt \int d^3x  \int d^3x'
  \int_{-\infty}^{\infty} dt' \times
  %\\\nonumber &\qquad
  V_\text{ext}(\mathbf{x},t)\frac{\partial
    \chi_{nn}^R(\mathbf{x},\mathbf{x}',t-t')}{\partial t}
  V_\text{ext}(\mathbf{x}',t')
  \\ \nonumber 
  &= - \frac{1}{\tau}  \int_0^{\tau}\!dt \int d^3x  \int d^3x'
  \int_{-\infty}^{\infty}\! dt' \int \frac{d^3q}{(2 \pi)^3}  \int
  \frac{d^3q'}{(2 \pi)^3} 
  %\times
  %\\\nonumber 
  %&\qquad
  \sum_{\mathbf{G}} \int \frac{d^3 Q}{(2
    \pi)^3}\int_{-\infty}^{+\infty} \frac{d\omega}{2 \pi}
  e^{-i \mathbf{q}\cdot \mathbf{x}_0} e^{i \mathbf{q}'\cdot \mathbf{x}_0}
  e^{i (\mathbf{q+Q})\cdot \mathbf{x}} \times
  \\ \nonumber 
  &
  \qquad  e^{-i(\mathbf{q'+Q+G})\cdot \mathbf{x}'} e^{-i t(\omega + \mathbf{q} \cdot
    \mathbf{v}_\text{SL})} e^{i t'(\omega + \mathbf{q}' \cdot
    \mathbf{v}_\text{SL})}(-i\omega)
%\times
%\\\nonumber &
%\qquad
  \,\chi_{nn}^R(\mathbf{Q},\mathbf{Q+G},\omega)\,
  \tilde{V}(| \mathbf{q}|) \tilde{V}(| \mathbf{q}'|) \, .
\end{align}
The integrations over $\mathbf{x}$, $\mathbf{x}'$, and $t'$ yield
Dirac-$\delta$ functions over $\mathbf{q}$, $\mathbf{q}'$ and $\omega$,
respectively.
Accordingly:
\begin{align}
  \bar{W} =&  \frac{i}{\tau}  \int_0^{\tau} dt 
  \sum_{\mathbf{G}} 
  \, e^{-i \mathbf{G}\cdot \mathbf{x}_0}
  \, e^{-i t \mathbf{G} \cdot \mathbf{v}_\text{SL}} 
  \int \frac{d^3 Q}{(2 \pi)^3}
  %\times
  %\\\nonumber &
  (\mathbf{Q+G})\cdot \mathbf{v}_\text{SL} \, 
  \chi_{nn}^R(\mathbf{Q},\mathbf{Q+G},(\mathbf{Q+G})\cdot \mathbf{v}_\text{SL}) \, 
  %\times
  %\\\nonumber &
  \tilde{V}(| \mathbf{Q}|) \, \tilde{V}(|\mathbf{Q+G}|)
  \nonumber \\ \label{power:eq}
  &= i  \sum_{\mathbf{G}_{\perp}} e^{-i \mathbf{G_{\perp}}\cdot
    \mathbf{x}_0} \int \frac{d^3 Q}{(2 \pi)^3}\,
  \mathbf{Q}\cdot \mathbf{v}_\text{SL} \, 
  \chi_{nn}^R(\mathbf{Q},\mathbf{Q}+\mathbf{G}_{\perp},\mathbf{Q}\cdot
  \mathbf{v}_\text{SL}) \, \tilde{V}(| \mathbf{Q}|) \, 
  \tilde{V}(|\mathbf{Q}+\mathbf{G}_{\perp}|)
  \,,
\end{align}
\end{widetext}
where we used the fact that the $t$ integration $\int_0^{\tau}dt\, e^{-i t \mathbf{G} \cdot \mathbf{v}_\text{SL}}$ vanishes 
whenever $\mathbf{G} \cdot \mathbf{v}_\text{SL} \neq 0$, and that 
\begin{equation}\label{deltaGparallel}
  \int_0^{\tau}dt\, e^{-i t \mathbf{G} \cdot \mathbf{v}_\text{SL}}
  = \tau \delta_{G_{\parallel},0}
  \,,
\end{equation}
$G_{\parallel}$ being the component of $\mathbf{G}$ parallel of the
sliding velocity, here simply the $x$ component of $\mathbf{G}$.
This leads to restricting the $\mathbf{G}$ summation to $\mathbf{G}_{\perp}$, the component of $\mathbf{G}$ perpendicular
to $\mathbf{v}_\text{SL}$, i.e.\ in the $yz$ plane.
In 2D, the integration becomes $d^2Q/(2\pi)^2$, and $\mathbf{G}_{\perp}$ coincides with the sole $G_y$ component.
In the expression for $\bar{W}$, Eq.~\eqref{power:eq}, the only factor which depends on the initial position
is $e^{-i \mathbf{G}_{\perp}\cdot \mathbf{x}_0}$.
Clearly, the parallel ($x$) component of the initial position is
irrelevant, since it corresponds to a trivial shift in time $t$, and the
dissipated power is averaged over $t$ anyway.
For the adopted symmetrically-located channeling trajectories
(see Sect.~\ref{model:sec}), this $\mathbf{x}_0$-related phase factor acquires
either of the following simple expressions:
\begin{align}\label{phase1:eq}
  e^{-i \mathbf{G_{\perp}}\cdot \mathbf{x}_0} &= (-1)^{l_y+l_z}
  &\text{if } \mathbf{x}_0&=\frac{a}{2}(\mathbf e_y+\mathbf e_z)
  \\\label{phase2:eq}
  e^{-i \mathbf{G_{\perp}}\cdot \mathbf{x}_0} &= (-1)^{l_y}    
  &\text{if } \mathbf{x}_0&=\frac{a}{2}\mathbf e_y \, .
\end{align}
Equation~\eqref{phase1:eq} is relevant in the 3D geometry only, while
Eq.~\eqref{phase2:eq} holds in both 3D and 2D.

In Appendix~\ref{symm:sec}
we analyze the parity of the integrand of Eq.~\eqref{power:eq}, arriving at Eq.~\eqref{Xisymmfin:eq}. 
That result proves that the real part of 
$\chi_{nn}^R(\mathbf Q,\mathbf Q+\mathbf G_\perp,\mathbf Q \cdot \mathbf
  v_\text{SL})\tilde{V} (\left| \mathbf Q + \mathbf G_\perp \right|)
  +
  \chi_{nn}^R(\mathbf Q,\mathbf Q-\mathbf G_\perp,\mathbf Q \cdot \mathbf
  v_\text{SL}) \tilde{V} (\left| \mathbf Q - \mathbf G_\perp \right|) 
$
is even under the transformation $\mathbf{Q} \to -\mathbf{Q}$, while its imaginary part is odd.

The full integrand of Eq.~\eqref{power:eq} can then be written as the product of a
$\mathbf{Q}$-odd factor, namely $\mathbf{Q}\cdot \mathbf{v}_\text{SL}$,
times the term whose symmetry properties have just been discussed.
As the $\mathbf{Q}$ integration is carried out over an even domain, only
the imaginary part of the retarded LR function contributes.
The dissipated power and therefore the friction force, see Eq.~\eqref{WFtimesV}, can then be expressed as:
\rem{
\begin{align}\label{power_imm:eq}
  \bar{W} =& - \sum_{\mathbf{G_{\perp}}} e^{-i \mathbf{G_{\perp}}\cdot \mathbf{x}_0}
  \int \frac{d^3 Q}{(2 \pi)^3}  \,\mathbf{Q}\cdot \mathbf{v}_\text{SL} \times
\\\nonumber&
  \IM\chi_{nn}^R(\mathbf{Q},\mathbf{Q+G_{\perp}},\mathbf{Q}\cdot \mathbf{v}_\text{SL})
  \tilde{V}(| \mathbf{Q}|) \tilde{V}(| \mathbf{Q+G_{\perp}}|).
\end{align}
The relation \eqref{WFtimesV} between the averaged dissipative power and
the friction force allows us to obtain
}
\begin{widetext}
\begin{align}\label{friction_imm:eq}
F(v_\text{SL}) =& \frac{\bar{W}}{v_\text{SL}} =
  - \sum_{\mathbf{G_{\perp}}} e^{-i \mathbf{G_{\perp}}\cdot \mathbf{x}_0}
  \int \frac{d^3 Q}{(2 \pi)^3}     \,
  \mathbf{Q}\cdot \hat{\mathbf{v}}_\text{SL} \, 
  %\times \\\nonumber &
  \IM\chi_{nn}^R(\mathbf{Q},\mathbf{Q+G_{\perp}},\mathbf{Q} \cdot
  \mathbf{v}_\text{SL}) \, 
 %\times  \\\nonumber &
  \tilde{V}(|\mathbf{Q}|) \, \tilde{V}(|\mathbf{Q+G_{\perp}}|)
 \,, 
\end{align}
where $\hat{\mathbf{v}}_\text{SL}= \mathbf{v}_\text{SL}/
|\mathbf{v}_\text{SL}|$ is the velocity unit vector.
\end{widetext}
As both $\tilde{V}$ factors in Eq.~\eqref{friction_imm:eq} are proportional
to the coupling strength $\varepsilon$, in the weak-coupling regime addressed
here the friction force is evidently a second-order, i.e.\ $\propto
\varepsilon^2$, effect.

In the 1D problem of Ref.~\cite{Panizon18}, the determination of the
imaginary part of the density-density response function relied on the
dynamic structure factor, which however led to an imprecise
expression \cite{Panizon24erratum}.
Therefore, here we stick to the imaginary part of this response function,
which we evaluate for the 2D or 3D harmonic crystal in
Appendix~\ref{correlation:app}.

\section{Explicit expression for friction}\label{Force:sec}

Starting from Eq.~\eqref{friction_imm:eq}, we come now to derive and
discuss an analytical expression for the friction force, assuming first a
conservative crystal, and then a dissipative one characterized by a finite
phonon lifetime.

\subsection{Conservative crystal}\label{conservative:sec}

We adopt the one-phonon approximation to the density-density response
function, Eq.~\eqref{chi_immnond:eq}, compute it for $\omega =
\mathbf{Q}\cdot \mathbf{v}_\text{SL}$, and substitute it into
Eq.~\eqref{friction_imm:eq}, obtaining:
\begin{widetext}
\begin{align}\label{force_nonrestrict:eq}
  F(v_\text{SL})&=
  \frac{\pi}{2 m a^3} \sum_{\mathbf G_\perp} 
  e^{-i \mathbf x_0 \cdot \mathbf G_\perp} \! \! 
  \int \frac{d^3Q}{(2\pi)^3}
  \mathbf{Q} \cdot  \hat{\mathbf{v}}_\text{SL}
  %\times \\\nonumber &
  \tilde{V} (\left|\mathbf{Q} \right|) \, 
  \tilde{V} (\left| \mathbf{Q} + \mathbf{G}_\perp \right|)
  e^{-W(\mathbf{Q})} e^{-W(\mathbf{Q}+\mathbf{G}_\perp)}
  \times \\\nonumber & \qquad
  \sum_{\lambda} \frac{
    \mathbf Q \cdot \boldepsilon_{\lambda}(\mathbf{Q}) \,
    (\mathbf Q +\mathbf G_\perp) \cdot \boldepsilon_{\lambda}(\mathbf{Q})
  }{ \omega_{\lambda}(\mathbf{Q})}
  %\times \\\nonumber &
  \Big[\delta(\mathbf{Q} \cdot \mathbf{v}_\text{SL} -
    \omega_{\lambda}(\mathbf{Q})) - \delta(\mathbf{Q} \cdot
    \mathbf{v}_\text{SL} + \omega_{\lambda}(\mathbf{Q})) \Big]
  \,,
\end{align}
\end{widetext}
where $e^{-W(\mathbf{Q})}$ and $e^{-W(\mathbf{Q}+\mathbf{G}_\perp)}$ are
Debye-Waller factors, see Eq.~\eqref{debyewaller:eq}.
This expression is directly suitable for the calculation of friction.
However it is conveniently simplified by taking advantage of the symmetry
of the polarization vectors of the phonons of the simple-cubic lattice, as
discussed in Appendix~\ref{lattice:app}, and summarised by
Eq.~\eqref{epsilon_refl_symm}.
With this information we can determine the parity of the integrand of
Eq.~\eqref{force_nonrestrict:eq} against reflections of each Cartesian
component of the vector $Q$.

Consider first the $\mathbf Q$ components perpendicular to the slider
velocity: $Q_y$ and $Q_z$.
The reflections $Q_y \to -Q_y$ and $Q_z \to -Q_z$ are equivalent by
symmetry, thus we discuss the first one only.
The factors $\tilde{V} (\left| \mathbf Q \right|)$, $e^{-W(\mathbf{Q})}$
and $\omega_{\lambda}(\mathbf{Q})$ are certainly even under this
reflection.
The remaining part of Eq.~\eqref{force_nonrestrict:eq} is:
\begin{align}\nonumber
  \mathcal{T} =
  \sum_{\mathbf G_\perp} e^{-i \mathbf x_0 \cdot \mathbf G_\perp} &
  \int \frac{d^3Q}{(2\pi)^3} 
  \tilde{V} (\left| \mathbf Q + \mathbf G_\perp \right|)\,
  e^{-W(\mathbf{Q}+\mathbf{G}_\perp)}
  \times \\ \nonumber 
  & \sum_{\lambda}  \mathbf Q \cdot \boldepsilon_{\lambda}(\mathbf Q) \,
  (\mathbf Q +\mathbf G_\perp) \cdot
  \boldepsilon_{\lambda}(\mathbf{Q})
  \,.
\end{align}
We divide the
$\mathbf{G}_\perp = \frac{2 \pi}{a} (0,l_y,l_z)$ vectors into 3 sets:
$\mathcal{G}_\perp^{0}$ including the $l_y = 0$ terms,
$\mathcal{G}_\perp^{+}$ including those with $l_y > 0$, and
$\mathcal{G}_\perp^{-}$ those with $l_y < 0$.
Therefore:
\begin{align}\nonumber
  \mathcal{T} =&
  \left[
    \sum_{G_\perp\in \mathcal{G}_\perp^0}+
    \sum_{G_\perp\in \mathcal{G}_\perp^+}+
    \sum_{G_\perp\in \mathcal{G}_\perp^-}
    \right]
    e^{-i \mathbf x_0 \cdot \mathbf G_\perp}\times
  \\\nonumber &
  \int \frac{d^3Q}{(2\pi)^3} \tilde{V} (\left| \mathbf Q +
  \mathbf G_\perp \right|) e^{-W(\mathbf{Q}+\mathbf{G}_\perp)}
  \\\label{termT:eq} 
  &
  \sum_{\lambda}  \mathbf Q \cdot 
  \boldepsilon_{\lambda}(\mathbf Q)  (\mathbf Q +\mathbf G_\perp) 
  \cdot
  \boldepsilon_{\lambda}(\mathbf{Q})
  \,.
\end{align}
The $\mathcal{G}_\perp^{0}$ terms are clearly even under the $Q_y \to -Q_y$
transformation.
Thanks to the discussed symmetry properties of the polarization vectors
$\boldepsilon_{\lambda}(\mathbf{Q})$, when we change $Q_y \to -Q_y$ the
$\mathcal{G}_\perp^+$ summation becomes equal to the $\mathcal{G}_\perp^-$
sum and vice versa.
We conclude that the sum of Eq.~\eqref{termT:eq}, and therefore
Eq.~\eqref{force_nonrestrict:eq} is even with respect to reflections of
$\mathbf Q$ in the directions perpendicular to the slider velocity.

As for the dependence on the parallel component $Q_x \equiv \mathbf{Q}
\cdot \hat{\mathbf{v}}_\text{SL}$, the $\mathcal{T}$ factor is certainly
even under $Q_x \to -Q_x$.
In Eq.~\eqref{force_nonrestrict:eq}, $Q_x$ affects the factor
\begin{equation}
  Q_x \left[ \delta(Q_x v_\text{SL} - \omega_{\lambda}(\mathbf{Q})) -
    \delta(Q_x v_\text{SL} + \omega_{\lambda}(\mathbf{Q})) \right] \, ,
\end{equation}
as well.
We see that this factor is also even under $Q_x \to -Q_x$.

Thanks to these symmetry considerations, it is possible to restrict the
momentum-space integration to the subdomain $\Omega$ defined by $(Q_x\geq
0 , Q_y \geq 0,Q_z \geq 0)$, provided that a factor 8 is included to
account for the omitted integration sectors.
For positive $v_\text{SL}$, the anti-resonant condition $\delta(Q_x
v_\text{SL} + \omega_{\lambda}(\mathbf{Q}))$ never occurs in the $\Omega$
subdomain, which allows us to simplify Eq.~\eqref{force_nonrestrict:eq}
to:
\begin{widetext}
\begin{align}
  F(v_\text{SL}) &= \frac{4\pi}{ m a^3} \sum_{\mathbf G_\perp} 
  e^{-i \mathbf x_0 \cdot \mathbf G_\perp} 
  \int_{\Omega} \frac{d^3Q}{(2\pi)^3} \, Q_x
  %\times \nonumber \\&
  \tilde{V} (\left| \mathbf Q \right|) \, 
  \tilde{V} (\left| \mathbf Q + \mathbf G_\perp \right|)\,
  e^{-W(\mathbf{Q})-W(\mathbf{Q}+\mathbf{G}_\perp)}
  \times \nonumber \\ \label{friction_nondissfinal:eq}
  & \hspace{46mm} \sum_{\lambda} \frac{\mathbf Q \cdot
  \boldepsilon_{\lambda}(\mathbf Q)  \, (\mathbf Q +\mathbf G_\perp)
  \cdot \boldepsilon_{\lambda}(\mathbf{Q})}{ \omega_{\lambda}(\mathbf{Q})} \, 
  \delta(Q_x v_\text{SL} - \omega_{\lambda}(\mathbf{Q}))
  \,.
\end{align}
\end{widetext}
This is one of the main analytic results of the present paper: an explicit
expression for the friction force as a function of the slider velocity for
a classical particle channelling at a constant velocity through a
conservative simple-cubic harmonic crystal.
For the 2D version of Eq.~\eqref{friction_nondissfinal:eq}, we defer to
Sect.~\ref{2D:sec}.
Equation~\eqref{friction_nondissfinal:eq} expresses the friction force as a
summation over the $\mathbf G$ vectors perpendicular to the sliding
direction of $\mathbf Q$-space integrals of products of (i) Fourier
components of the slider-crystal interaction, (ii) Debye-Waller factors,
(iii) projections of the phonon polarization vectors in the $\mathbf Q$ and
$\mathbf Q+\mathbf G$ directions, the ratio
$Q_x/\omega_{\lambda}(\mathbf{Q})$, and (iv) a Dirac delta imposing the
resonance 
condition
\begin{align}\label{resonance:eq}
   \omega_{\lambda}(\mathbf{Q}) = \mathbf{Q}\cdot \mathbf{v}_\text{SL}
  \,,
\end{align}
namely that the product $\mathbf{Q}\cdot \mathbf{v}_\text{SL} \equiv Q_x
v_\text{SL}$ matches the phonon frequency at the same $\mathbf Q$ point.

This Dirac-delta allows us to restrict
the $\mathbf{Q}$-integration using the following identity:
\begin{equation}\label{delta:eq}
  \int_\Omega d^3Q \,g(\mathbf{Q}) \delta(f(\mathbf{Q}))
  = \int_{S^\Omega} d^2Q \,
  \frac{g(\mathbf{Q})}{|\boldsymbol{\nabla} f(\mathbf{Q}) |}
  \,,
\end{equation}
where $S^\Omega$ is the 2D manifold which solves the equation
$f(\mathbf{Q})=0$ in the $\Omega$ subset of the $\mathbf Q$ space.
In our Eq.~\eqref{friction_nondissfinal:eq}, $f(\mathbf{Q})=\mathbf{Q}\cdot
\mathbf{v}_\text{SL} - \omega_{\lambda}(\mathbf{Q})$.
As a consequence, 
$S^\Omega_\lambda$ is the manifold of solutions of Eq.~\eqref{resonance:eq}
for phonon branch $\lambda$;
additionally, the gradient $\boldsymbol{\nabla}
f(\mathbf{Q})= \mathbf{v}_\text{SL} -\mathbf{v}_{\lambda}(\mathbf{Q})$,
where $\mathbf{v}_\lambda(\mathbf{Q}) \equiv \boldsymbol{\nabla}_{\mathbf Q}
\omega_\lambda(\mathbf{Q})$ is the group velocity associated to phonon
branch $\lambda$.
Accordingly, the implementation of the recipe \eqref{delta:eq} transforms Eq.~\eqref{friction_nondissfinal:eq} to
\begin{widetext}
\begin{align} \label{force_delta:eq} %\nonumber
  F(v_\text{SL})& = \frac{4\pi}{ m a^3} \sum_{\mathbf G_\perp} 
  e^{-i \mathbf x_0 \cdot \mathbf G_\perp} \sum_{\lambda}
  \int_{S^\Omega_\lambda} \frac{d^2Q}{(2\pi)^3}
  \,Q_x
%  \times \\\nonumber&
  \tilde{V} (\left| \mathbf Q \right|) \tilde{V}
  (\left| \mathbf Q + \mathbf G_\perp \right|)\,
  e^{-W(\mathbf{Q})-W(\mathbf{Q}+\mathbf{G}_\perp)} 
%  \times \\ \label{force_delta:eq} &
  \frac{ \mathbf Q \cdot \boldepsilon_{\lambda}(\mathbf Q) \,
    (\mathbf Q +\mathbf G_\perp) \cdot
    \boldepsilon_{\lambda}(\mathbf{Q}) }
    {\omega_{\lambda}(\mathbf{Q})\,| \mathbf{v}_\text{SL} -
      \mathbf{v}_\lambda(\mathbf{Q})|}
  \,.
\end{align}
\end{widetext}
As we lack analytic formulas for the 3D dispersion relations $\omega_\lambda(\mathbf{Q})$,
we cannot express the manifolds $S^\Omega_\lambda$ explicitly, and therefore we
do not see how to further simplify Eq.~\eqref{force_delta:eq}.

Nevertheless, this formulation is highly instructive and useful as it stands.
Equation~\eqref{force_delta:eq} indicates that (i) friction comes from the
additive contribution of the three phonon branches; (ii) at any given
velocity $\mathbf{v}_\text{SL}$, friction is determined uniquely by a well-defined
set of phonon modes, namely those that lie on the $S^\Omega_\lambda$ surfaces in
$Q$ space; (iii) especially large friction is expected from the
contribution to the integral of phonons satisfying the ``shock-wave''
%\textcolor{blue}{Erio: I would call it  "velocity matching", but  maybe shock wave is more shocking..:-))) }
condition
\begin{equation}\label{shock-wave:eq}
  \mathbf{v}_\text{SL} =\mathbf{v}_{\lambda}(\mathbf{Q}) 
  \,,
\end{equation}
that makes the denominator vanish.
We shall return to this shock-wave condition below.

Note importantly that, regardless of the sliding speed $v_\text{SL}$, for each
phonon branch $\lambda$, there certainly exists some nonempty $S^\Omega_\lambda$
surface.
The reason is that the condition \eqref{resonance:eq} for $\mathbf{Q}$
sitting on the $S^\Omega_\lambda$ surface can certainly be met at least at
sufficiently remote (i.e.\ large $|Q_x|$ or large $|Q_{y/z}|$) regions of
$\mathbf Q$ space.
As a result, for any speed $v_\text{SL}$ the conservative crystal keeps resonant phonon dissipation
channels open, resulting in nonzero friction.
This finding is qualitatively different from what was found for the 1D
model \cite{Panizon18}: in 1D, when $v_\text{SL}$ exceeds the speed of
sound, no energy transfer to the conservative chain is possible, 
%with the peculiar result of 
leading instead to
vanishing friction for supersonic sliding.
 
\subsection{Dissipative crystal}\label{dissipative:sec}

We proceed now to verify how a weak decay of the phonons in the crystal affects friction.
By substituting the one-phonon approximation of the density-density
response function of the dissipative crystal, Eq.~\eqref{chi_immd:eq}, into
Eq.~\eqref{friction_imm:eq} we obtain:
\begin{widetext}
\begin{align} \nonumber
  F(v_\text{SL}) &=
  \frac{\pi}{2 m a^3} \sum_{\mathbf G_\perp} 
  e^{-i \mathbf x_0 \cdot \mathbf G_\perp} 
  \int \frac{d^3Q}{(2\pi)^3}
  \, Q_x
%  \\\nonumber  &\times
  \tilde{V} (\left| \mathbf Q \right|) \, 
  \tilde{V} (\left| \mathbf Q + \mathbf G_\perp \right|) \,  
  e^{-W(\mathbf{Q})-W(\mathbf{Q}+\mathbf{G}_\perp)} \times 
  \\ \label{force_nonrestrictdiss:eq}  & \hspace{45mm} 
  \sum_{\lambda}
  \mathbf{Q} \cdot \boldepsilon_{\lambda}(\mathbf{Q}) \,
    (\mathbf{Q} +\mathbf{G}_\perp) \cdot
    \boldepsilon_{\lambda}(\mathbf{Q}) 
%  \\ &\times
  \,\mathcal{L}(\mathbf{Q}, v_\text{SL}, \gamma)
  \,, 
\end{align}
with
\begin{align}
\label{doubleLorenz:eq} 
  \mathcal{L}(\mathbf{Q}, v_\text{SL}, \gamma) &=
  [\omega_{\lambda}(\mathbf{Q})]^{-1}\left[
  \frac{\frac{\gamma}{2\pi}}{(Q_x v_\text{SL} -
    \omega_{\lambda}(\mathbf{Q}))^2 + (\frac{\gamma}{2})^2}
%  \\ &\qquad 
  -
  \frac{\frac{\gamma}{2\pi}}{(Q_x v_\text{SL} +
    \omega_{\lambda}(\mathbf{Q}))^2 + (\frac{\gamma}{2})^2}
    \right]
  \\ \label{doubleLorenz_final:eq} % \label{doubleLorenz_again:eq} 
  &= \frac{\gamma}{2\pi}\,
  \frac{4 Q_x v_\text{SL}}
  {[(Q_x v_\text{SL} -
    \omega_{\lambda}(\mathbf{Q}))^2 + (\frac{\gamma}{2})^2] \, [(Q_x v_\text{SL} +
    \omega_{\lambda}(\mathbf{Q}))^2 + (\frac{\gamma}{2})^2]}
%    \\=&
%  \frac{\gamma}{2\pi}\,
%  \frac{4 Q_x v_\text{SL}}
%  {\left((Q_x v_\text{SL})^2+
%  \omega_{\lambda}^2(\mathbf{Q})+
%  (\gamma/2)^2
%  \right)^2
%  -(2 Q_x v_\text{SL}
%  \omega_{\lambda}(\mathbf{Q}))^2}
  \,.
\end{align}

\end{widetext}
The formulation \eqref{doubleLorenz_final:eq} explicitly clarifies that no singularity arises at the $\mathbf{Q}$ points where $\omega_{\lambda}(\mathbf{Q})$ vanishes, and was anticipated in Eq.~\eqref{friction_dissfinal:eq} in the Introduction.
Dimensionally, the quantity $\mathcal{L}(\mathbf{Q}, v_\text{SL}, \gamma)$ is a squared time and combines the double Lorentzian terms with the inverse phonon-frequency dependence in the response function of Eq.~\eqref{chi_immd:eq}.
The integrand function in Eq.~\eqref{force_nonrestrictdiss:eq} exhibits the same symmetries against reflections in the
$Q_y$ and $Q_z$ components perpendicular to $v_\text{SL}$, as discussed
above for the conservative crystal.
As for the $Q_x \to -Q_x$ reflection, all factors are symmetric, except for
$Q_x$ in the first line and the function $\mathcal{L}(\mathbf{Q},
v_\text{SL}, \gamma)$: both these terms change sign under this reflection.
As a result, the integrand function is fully symmetric against all
$Q_\alpha \to -Q_\alpha$ reflections.
Thanks to these symmetries, we can reduce the momentum-space integration to the one-eight sub-domain $\Omega$ defined previously, obtaining Eq.~\eqref{friction_dissfinal:eq}, which is the second fundamental result of the present paper.

The integral in Eq.~\eqref{friction_dissfinal:eq} extending over one eight of
the reciprocal space can be evaluated numerically, as discussed in the
next Section.
While for the conservative
crystal Eq.~\eqref{force_delta:eq} predicts that only the phonon modes whose $\mathbf Q$ satisfies the resonance condition \eqref{resonance:eq} contribute to friction, the main novelty for the dissipative crystal is that
Eq.~\eqref{friction_dissfinal:eq} gathers friction contributions from {\em all} $\mathbf Q$ points.
Certainly, the distribution $\mathcal{L}(\mathbf{Q},
v_\text{SL}, \gamma)$, Eq.~\eqref{doubleLorenz:eq}, gives more weight to contributions from the $\mathbf Q$-space regions near the $S^\Omega_\lambda$
surfaces satisfying Eq.~\eqref{resonance:eq} than to more remote regions.
In the limit $\gamma\to0$ the antiresonant term in
Eq.~\eqref{doubleLorenz:eq} vanishes, and the integral focuses sharply onto
the $S^\Omega_\lambda$ surfaces, recovering the result \eqref{force_delta:eq}
valid for infinite-lifetime phonons.

Equation~\eqref{friction_dissfinal:eq} makes explicit predictions for the friction force originated from weak phonon emission of a point slider crossing a weakly damped harmonic crystal at arbitrary velocity.
The resulting friction force can be compared directly against numerical simulations, and in principle even against actual channeling experiments.

\section{Evaluation of friction}\label{3D:sec}

In this section we report the results of the 
%numerical
evaluation of the main expression of the present work, Eq.~\eqref{friction_dissfinal:eq}, for specific choices of the parameters of the 3D model.
We focus on the dissipative crystal for 2 very practical reasons:
(i) The alternative evaluation of Eq.~\eqref{force_delta:eq} would require
an accurate numerical determination of the $S^\Omega_\lambda$ surfaces, their
parameterization, and the implementation of an appropriate
surface-integration algorithm.
While this is certainly feasible, it is a rather cumbersome task, highly
prone to numerical instabilities and implementation mistakes.
(ii) It would be practically very difficult, or outright impossible, to compare the resulting friction force with a dissipation-free simulated MD mechanical model.
Indeed sliding in a finite-size conservative crystal would progressively
heat it up, with temperature and thus friction drifting systematically in time, thus making a steady state unreachable.

For definiteness, we focus on the $T \to 0$ limit, thus
%simulating the crystal as classical, but ignoring 
leaving any
%all weak  moderate 
temperature effects out.
In this limit the mean phonon numbers $n_{\lambda}({\mathbf k})$ vanish.
Consequently the expression Eq.~\eqref{debyewaller:eq} for the Debye-Waller
factor simplifies to
\begin{equation}\label{debye_T0:eq}
  W(\mathbf{Q}) = \frac{1}{2N} \sum_{\mathbf{k},\lambda} | \mathbf{Q} \cdot
   \boldepsilon_{\lambda}(\mathbf{k}) |^2 \,
  \frac{\hbar}{2m\omega_{\lambda}(\mathbf{k})}
  \,,
\end{equation}
accounting for zero-point motion.
The final fraction in Eq.~\eqref{debye_T0:eq} represents the square of the characteristic length of the quantum harmonic oscillator labeled by $\mathbf{k},\lambda$: in practice the squared amplitude of the spatial quantum fluctuations of that phonon mode.
For typical phonon frequencies this amplitude is quite small, $\approx 4$~pm for the crystal parameters of Table~\ref{uofm:tab}.
Accordingly, for the relatively short $|{\mathbf Q}|$ vectors that dominate the integral in Eq.~\eqref{friction_dissfinal:eq}, most individual terms in the summation of Eq.~\eqref{debye_T0:eq} are quite small too.
As a result unity is a fair approximation for both Debye-Waller factors: $\exp(-W(\mathbf{Q}))\simeq 1$ and
$\exp(-W(\mathbf{Q}+\mathbf{G}_\perp))\simeq 1$.

In contrast, for large enough $|\mathbf{Q}|$, this approximation fails, especially for soft phonon modes, which are associated with wider fluctuations, leading to Debye-Waller factors deviating from unity.
In practice, in this same large-$Q$ region, having assumed a well-behaved slider-crystal potential, the decay of its Fourier transform $\tilde{V}(Q)$, see Eq.~\eqref{pot_fourier:eq}
and Fig.~\ref{potq:fig}, makes these large-$|\mathbf{Q}|$ contributions to
friction, and the associated deviations of the Debye-Waller factors from
unity, negligible anyway.
If one wished to address a stiffer potential, characterized by slowly decaying Fourier components $\tilde{V}(Q)$, then precisely the large-$|\mathbf{Q}|$ decay of the Debye-Waller factors will ensure the convergence of the $Q$ integration even in such a situation.

According to this analysis, in the following we substitute unity in place of the Debye-Waller factors of Eq.~\eqref{friction_dissfinal:eq}.
Observe that in this low-temperature gentle-interaction regime where the
Debye-Waller factors approach unity, Planck's constant $\hbar$ disappears from
the friction expression altogether, indicating that no quantum effects should be relevant %\textcolor{blue}{
in the present high-velocity, essentially harmonic-crystal conditions -- the damping $\gamma$ assumed to be accordingly small.

A larger damping rate $\gamma$ makes the Lorentzian weight a smoother function of $\mathbf{Q}$, which in turn makes the convergence of the integration in Eq.~\eqref{friction_dissfinal:eq}
faster, as the $\mathbf{Q}$-space grid is refined.
In contrast, small $\gamma$ determines a rapidly varying integrand function
and, as observed above, concentrates most integration weight sharply around the
$S^\Omega_\lambda$ surfaces: a converged integral evaluation requires a very fine
$\mathbf{Q}$-space sampling, but comes with the bonus of providing finer
details of the velocity dependence of friction.
As a fair compromise, for the numerical evaluation of
Eq.~\eqref{friction_dissfinal:eq}, we adopt an intermediate damping rate
$\gamma=0.2\,(K/m)^{1/2}$ which makes the Lorentzian broadening factor
smooth enough for a feasible numerical integration.

\begin{figure} 
\centerline{
\includegraphics[width=\linewidth,clip=]{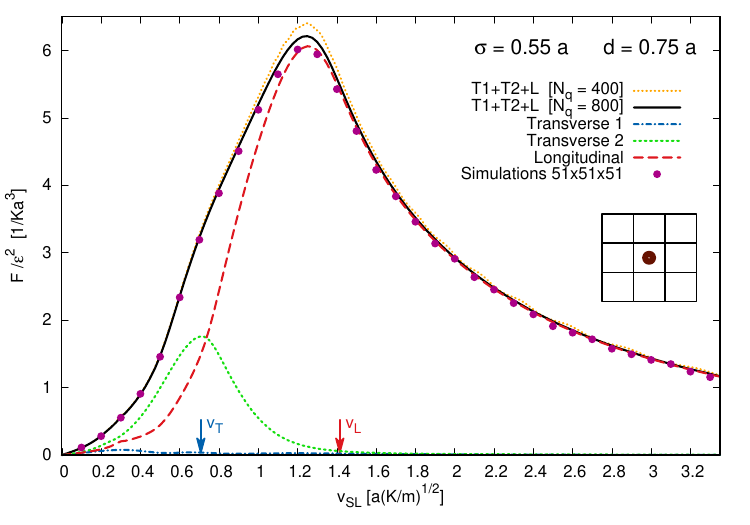}
}
\caption{\label{cube_center:fig}
Friction force as a function of the slider velocity, computed according to
Eq.~\eqref{friction_dissfinal:eq}, with the slider following the line
through the cube centers of the sc lattice, as determined by the initial
condition $\mathbf{x}_0=\frac a2(\mathbf{e}_y+\mathbf{e}_z)$.
The slider-crystal interaction potential, Eq.~\eqref{pot_fourier:eq}, is
parameterized by $\sigma=0.55 a$ and $d=0.75 \,a$.
The second-neighbor spring constant $K'=K/2$, and the crystal is damped by
a viscous friction with $\gamma=0.2\,(K/m)^{1/2}$.
Black solid line: the total friction.
Other curves: the contributions of individual polarization branches.
Points: friction evaluated by means of numerical simulations.
Arrows: the longitudinal (red) and transverse (blue) speeds of sound in the $(100)$ direction.
}
\end{figure}

The dispersion relations and the polarization vectors are obtained by
solving the secular equation \eqref{seculat:eq} numerically.
Regarding the interaction potential, we fix $\sigma = 0.55\,a$ and $d=0.75\,a$, as depicted in
Fig.~\ref{potq:fig}.
As the friction force is proportional to $\varepsilon^2$, in the following we
report friction divided by this perturbatively small quantity.

We construct a suite of python programs \cite{support_data_analytic3D}
for executing the integration and
summations indicated in Eq.~\eqref{friction_dissfinal:eq}.
We execute the $\mathbf{Q}$ integration over a wide cubic truncation of the region
$\Omega$, limiting each component to $0\leq Q_\alpha\leq Q^\text{max} = 2\pi/d_x$.
This trunction is consistent with a real-space spatial resolution $d_x$.
We adopt $d_x=0.15\,a$, a value that extends the
$\mathbf Q$-integration region up to  $Q^\text{max} \simeq 41.7\,a^{-1}$.
We have verified that, thanks to the rapid decay of the Fourier transform
of the potential (Fig.~\ref{potq:fig}) the results are weakly affected by
this cutoff.
For the integration, each component is sampled on a uniform grid of spacing $\Delta Q = Q^\text{max}/N_Q$, with $N_Q=800$.
This grid fineness parameter is the most delicate one for the convergence of the integration in Eq.~\eqref{friction_dissfinal:eq}.
As for the $\mathbf{G}_\perp$ summation, we include terms with $-l_\text{max}\leq l_{x,y} \leq l_\text{max}$, with $l_\text{max}=15$: we have verified that friction changes negligibly if we repeat the calculation with $l_\text{max}=10$.
The code pre-computes all
$v_\text{SL}$-independent relevant quantities in
Eq.~\eqref{friction_dissfinal:eq} at all $\mathbf Q$ grid points and
stores them in a set of files.
This trick allows us to then execute, for each $v_\text{SL}$, the discrete summation representing the $\mathbf Q$ integration much more quickly, since just the $\mathcal{L}(\mathbf{Q}, v_\text{SL}, \gamma)$
factor is left to evaluate.

\begin{figure} 
\centerline{
\includegraphics[width=\linewidth,clip=]{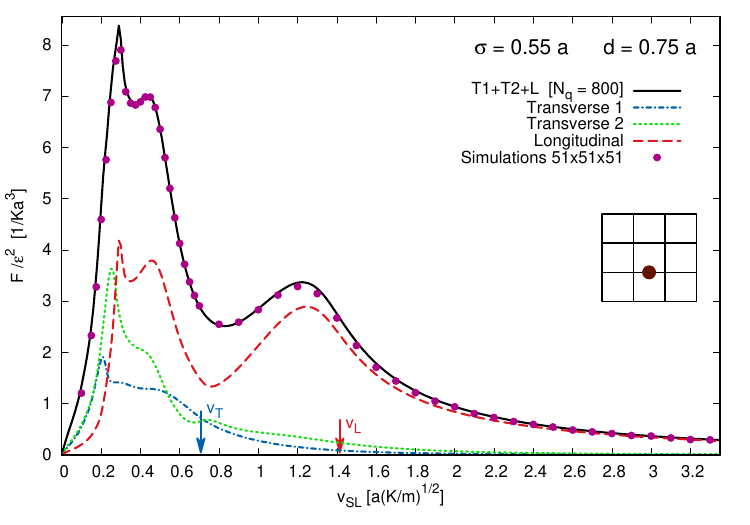}
}
\caption{\label{bond_center:fig}
Same as Fig.~\ref{cube_center:fig}, but with the slider following the line
midway between two rows of atoms, i.e.\ through nearest-neighbor bond centers of the sc lattice, as dictated by the initial condition $\mathbf{x}_0=\frac
a2\,\mathbf{e}_y$.
%
%For $v_\text{SL}$ smaller than $v_\text{T}$, observe strong contributions to friction from all polarizations, while above $v_\text{L}$ the longitudinal phonons dominate dissipation.
%
}
\end{figure}

Figures~\ref{cube_center:fig} and \ref{bond_center:fig} report the result
of the numerical integration of Eq.~\eqref{friction_dissfinal:eq} for the
channeling trajectories fixed by $\mathbf{x}_0=\frac a2 (\mathbf{e}_x +
\mathbf{e}_y)$ and $\mathbf{x}_0=\frac a2 \mathbf{e}_x$, respectively.
Observe that the velocity dependence is significantly different when the sliding particle follows the two alternative channeling lines, with friction peaking at different velocities.
In both figures, the contributions of the three phonon branches to friction are reported individually, showing that for small speed $v_\text{SL} < v_\text{T}$, longitudinal and transverse phonons contribute in similar proportions, while for larger speeds the longitudinal phonons dominate.

Importantly, both figures report the comparison of the analytic evaluation of the total friction (black solid curve) according to
Eq.~\eqref{friction_dissfinal:eq} with the outcome of MD simulations (pink dots), detailed in Appendix~\ref{MD:appendix}.
The agreement is striking.
Small deviations can be traced to the imperfect convergence of the integration grid.
The comparison of the $N_q=800$ curve with the less-converged $N_q=400$ curve (orange dotted curve of Fig.~\ref{cube_center:fig}) shows that the finer grid improves the agreement.

In comparing 
Fig.~\ref{bond_center:fig} to Fig.~\ref{cube_center:fig}, observe that
higher friction peaks are predicted when the slider follows
the bond-center path.
The reason is that the slider approaches the crystal particles more closely
(nearest approach distance $a/2$) than along the cube-center path, where
the nearest-approach distance equals $a/\sqrt{2}$.
At the closer approach distance, the repulsive region $r<\sigma=0.55\,a$ of
the interaction potential, Fig.~\ref{potx:fig}, is involved: as a result,
the slider-crystal collisions excite
phonons more effectively than the smoother attractive tail.

\subsection{An approximate expression}

Since the evaluation of Eq.~\eqref{friction_dissfinal:eq} requires a
relatively cumbersome integration, a simpler approximate expression for a
quick order-of-magnitude estimation of friction could be of some value.
We 
%derive this approximate formula from
simplify
Eq.~\eqref{friction_dissfinal:eq} by
(i) omitting the $\mathbf G_\perp$ and $\lambda$ summations;
(ii) replacing each $\mathbf Q$ component with $d^{-1}$
(thus $\left| \mathbf Q \right|\to \sqrt{3}/d$), estimating the region
where $\tilde{V}(\left| \mathbf Q \right|)$ 
peaks;
(iii) substituting the amplitude of the $d^3 Q$ integration with the volume $(2\pi/d)^3$ of the $\mathbf Q$-space region where $\tilde{V}(\left| \mathbf Q \right|)$ is nonnegligible, see Fig.~\ref{potq:fig}; 
(iv) replacing the polarization vectors $\boldepsilon_{\lambda}$ with unity;
(v) replacing the Debye-Waller exponentials with unity too:
\begin{align}\nonumber
  F(v_\text{SL})\approx\,&
  \frac{4\pi}{m a^3}
  \, \frac 1{d^3}
  \, d^{-1}
  \tilde{V}^2\!\left(\frac{\sqrt{3}}d\right)
  \!d^{-2}\,
  \mathcal{L}\!\left(d^{-1} \hat{\mathbf x}, v_\text{SL}, \gamma\right)
\\\label{Fricapprox1:eq} =\,&
\frac{4\pi}{m a^3}
\,
\frac{\tilde{V}^2\!\left(\frac{\sqrt{3}}d\right)}{d^6}\,
  \mathcal{L}\!\left(d^{-1} \hat{\mathbf x}, v_\text{SL}, \gamma\right)
.
\end{align}
To estimate the velocity-dependent $\mathcal{L}$ factor defined in Eq.~\eqref{doubleLorenz_final:eq}, we introduce a typical phonon frequency $\omega_\text{ph}\approx (K'/m)^{1/2}=(K/2m)^{1/2}$ and a scaled dimensionless slider velocity $w=v_\text{SL}/(d\,\omega_\text{ph})$.
In terms of these quantities, we estimate $\mathcal{L}$ as follows:
\begin{widetext}
\begin{align}\nonumber
\mathcal{L}\!\left(d^{-1} \hat{\mathbf x}, v_\text{SL}, \gamma\right) \approx\, &
\frac{\gamma}{2\pi}\,
  \frac{4 d^{-1} v_\text{SL}}
  {[(d^{-1} v_\text{SL} -
    \omega_\text{ph})^2 + (\frac{\gamma}{2})^2] \, [(d^{-1} v_\text{SL} +
    \omega_\text{ph})^2 + (\frac{\gamma}{2})^2]}
\\\label{Lapprox:eq} =\,&
\frac{2\gamma}{\pi\,\omega_\text{ph}^3}\,
  \frac{w}
  {[(w - 1)^2 + (\frac{\gamma}{2\omega_\text{ph}})^2] \, [(w +
    1)^2 + (\frac{\gamma}{2\omega_\text{ph}})^2]}
\,,
\end{align}
\end{widetext}
providing a typical order of magnitude.  The ratio
$\gamma/(2\omega_\text{ph})$ is inversely related to the quality factor of
a typical phonon oscillator.
The expression~\eqref{Lapprox:eq} provides a fair off-resonance ($w\ll1$ or
$w\gg1$) estimation for $\mathcal{L}$, that is not too bad even for speeds
matching a phonon resonance ($w\simeq 1$).
The product $v_s=a \,\omega_\text{ph}$ represents a typical sound velocity of the order of the speeds of sound evaluated in Appendix~\ref{lattice:app}.
In terms of these quantities, by combining Eqs.~\eqref{Fricapprox1:eq} and \eqref{Lapprox:eq}, the order of magnitude of the off-resonance friction force is
\begin{align}\nonumber
F\approx\,&
\frac{8 \, \gamma}{m v_s^3}\,
\frac{\tilde{V}^2\!\left(\frac{\sqrt{3}}{d}\right)}{d^6}\times
\\\label{frictionOoM}
&
  \frac{w}
  {[(w - 1)^2 + (\frac{\gamma}{2\omega_\text{ph}})^2] \, [(w +
    1)^2 + (\frac{\gamma}{2\omega_\text{ph}})^2]}
\\\nonumber \approx\, &
\frac{8\, \gamma}{m v_s^3}\,
\frac{\tilde{V}^2\!\left(\frac{\sqrt{3}}{d}\right)}{d^6}
\times
\begin{cases}
\frac w{\left[1+\left(\frac{\gamma}{2\omega_\text{ph}}\right)^2\right]^2} & \text{for } w\ll 1\\
\frac 1 {\left(\frac{\gamma}{2\omega_\text{ph}}\right)^2\left[2+\left(\frac{\gamma}{2\omega_\text{ph}}\right)^2\right]} & \text{for } w = 1\\
\frac 1{w^3} & \text{for } w\gg 1
\end{cases}
.
\end{align}
This expression is proportional to the square of the coupling energy $\varepsilon \propto\tilde{V}\!\left(\sqrt{3}/d\right)/d^3$, as expected of a linear-response result.

\begin{figure} 
\centerline{
\includegraphics[width=\linewidth,clip=]{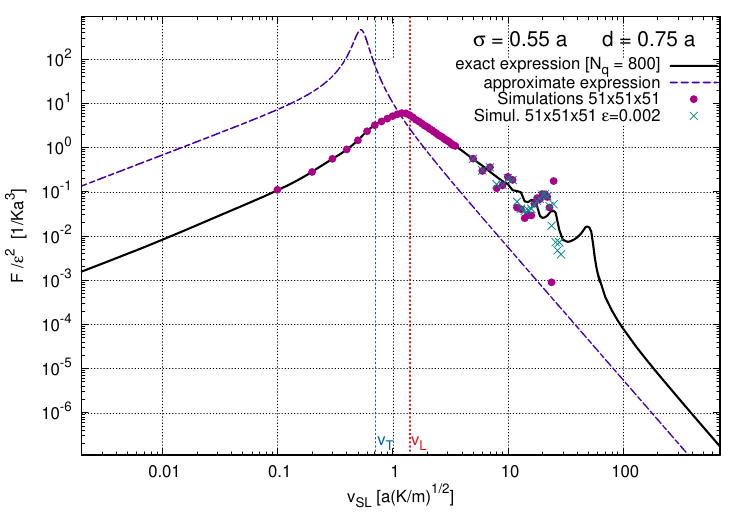}
}
\caption{\label{logfriction:fig}
A broad-range comparison of the total friction force as a function of the slider velocity, computed according to
Eq.~\eqref{friction_dissfinal:eq} (solid), with the approximate expression \eqref{frictionOoM} (dashed), for the same conditions and parameters as in Fig.~\ref{cube_center:fig}.
Points: friction evaluated by means of numerical simulations carried out with coupling $\varepsilon = 5\times 10^{-4}\,Ka^2$ (circles) or $\varepsilon = 0.002\,Ka^2$ (crosses).
%
% Possibly removable sentence:
Vertical dashed lines: the longitudinal (red) and transverse (blue) speeds of sound in the $(100)$ direction.
}
\end{figure}

Figure~\ref{logfriction:fig} compares the estimation of Eq.~\eqref{frictionOoM} with the full Eq.~\eqref{friction_dissfinal:eq} and with numerical simulations over a broad velocity range explored in logarithmic scale.
It is seen that the order-of-magnitude estimation Eq.~\eqref{frictionOoM} yields a weak friction $\propto v_\text{SL}$ for $v_\text{SL}\to 0$, 
a peaking friction related to the square of the typical quality factor of phonon oscillators at intermediate speed, and a rapidly decaying friction $\propto v_\text{SL}^{-3}$ in the large-$v_\text{SL}$ limit.
The rough estimation Eq.~\eqref{frictionOoM} is in fair qualitative agreement with the exact formula Eq.~\eqref{friction_dissfinal:eq}, with the correct power-law asymptotic dependences and a roughly positioned friction peak, but shows of course significant quantitative deviations, especially at low velocity.

Interestingly, the fact that the approximate formula depends on $v_\text{SL}$ through the dimensionless ratio $w=v_\text{SL}/(d\,\omega_\text{ph})$ suggests that relevant speed to which $v_\text{SL}$ should be compared is not the speed of sound $v_s$, but rather the combination $d\,\omega_\text{ph}$ of the characteristic length scale of the slider-crystal interaction and the typical vibration frequency of the crystal.

As a final note on Fig.~\ref{logfriction:fig},
the nearly perfect agreement of the full friction formula
\eqref{friction_dissfinal:eq} with the simulations in the velocity range
around the speeds of sound degrades somewhat at higher velocity.
This is due to multiple effects including: (i) the truncation of the
$\mathbf Q$ integration to a finite cutoff $Q^\text{max}$ affects the
wiggles of the theory curve;
(ii) the very low friction in this region makes its numeric evaluation
through MD simulations prone to roundoff error, which we reduce by
repeating the calculation with a larger coupling (cross points in
Fig.~\ref{logfriction:fig}).

\section{The 2D model}\label{2D:sec}

It is instructive to adapt the results obtained for the 3D crystal to a 2D square lattice.
First off, we turn all vector quantities into 2D vectors.
For the 2D crystal the integration domain $\Omega$ of
Eq.~\eqref{friction_nondissfinal:eq} is replaced by its 2D version, namely the quadrant $(Q_x\geq 0 , Q_y \geq 0)$.
The polarization index $\lambda$ takes only two values for the transverse and longitudinal phonon branches.
The sum over $\mathbf{G}_\perp$ involves just the $y$ component:
$$\mathbf{G}_\perp=(0,\frac{2\pi}{a}l_y)$$ associated to a single Miller index
$l_y$.
In adapting Eq.~\eqref{friction_nondissfinal:eq} to 2D, a power of $a$, a factor $2\pi$ at the denominator,
and a factor $2$ at the numerator originated by symmetry  must also be removed.
According to these observations, the 2D version of Eq.~\eqref{friction_nondissfinal:eq}, valid for infinitely-long-lived phonons, is
\begin{widetext}
\begin{align} \nonumber
F(v_\text{SL}) &= \frac{2\pi}{m a^2} \sum_{l_y} (-1)^{l_y}
  \int_{\Omega} \frac{d^2Q}{(2\pi)^2}
  Q_x 
  %\\ \nonumber & 
  \tilde{V} (\left| \mathbf Q \right|) \tilde{V}
  (\left| \mathbf Q + \mathbf G_y \right|)\, 
  e^{-W(\mathbf{Q})-W(\mathbf{Q}+\mathbf{G}_y)} \times
  \\ \label{friction_nondissfinal_2D:eq} %\nonumber
  & \hspace{45mm} \sum_{\lambda} \frac{\mathbf{Q} \cdot
    \boldepsilon_{\lambda}(\mathbf Q)
    \, (\mathbf Q +\mathbf G_y) \cdot
    \boldepsilon_{\lambda}(\mathbf{Q}) }
      { \omega_{\lambda}(\mathbf{Q})} \, 
  \delta(\mathbf{Q}\cdot \mathbf{v}_\text{SL} - \omega_{\lambda}(\mathbf{Q}))
  \,.
\end{align}
\end{widetext}
We can reformulate the 2D integral, using the following property of the Dirac delta:
%NICK: qui diceva 2D Dirca delta, ma era sbagliato! Questa e` una 1D Dirac delta, con come argomento una funzione di 2 variabili a valore scalare.
\begin{equation}
  \int_{\Omega} d^2Q \, g(\mathbf{Q}) \delta(f(\mathbf{Q}))= \int_{\ell^{\Omega}} dQ \, \frac{g(\mathbf{Q})}{|\boldsymbol{\nabla} f(\mathbf{Q}) |} \, ,
\end{equation}
where $\ell^{\Omega}$ is the curve consisting of the solutions of the equation
$f(\mathbf{Q})=0$ in the $\mathbf{Q}$-space domain $\Omega$.
Specifically for Eq.~\eqref{friction_nondissfinal_2D:eq} we take
\begin{equation}
f(\mathbf{Q})=\mathbf{Q}\cdot \mathbf{v}_\text{SL} -
\omega_{\lambda}(\mathbf{Q}) =Q_x v_\text{SL} -
\omega_{\lambda}(\mathbf{Q})
\,,
\end{equation}
and therefore
\begin{equation}
\boldsymbol{\nabla} f(\mathbf{Q})= \mathbf{v}_\text{SL}
-\mathbf{v}_\lambda(\mathbf{Q}) =
\left(
\begin{array}{c}
v_\text{SL}-v_{\lambda\,x}(\mathbf{Q})\\
-v_{\lambda\,y}(\mathbf{Q})
\end{array}
\right)
.
\end{equation}
Like in 3D, we introduce the group velocity of the phonons
$\mathbf{v}_\lambda(\mathbf{Q})= \boldsymbol{\nabla}
\omega_\lambda(\mathbf{Q})$, namely the gradient of their dispersion, with components
$v_{\lambda\,x/y}(\mathbf{Q})$.
For each polarization $\lambda$,
the solution of $Q_x \, v_\text{SL} = \omega_{\lambda}(\mathbf{Q})$  generates a specific curve $\ell_\lambda^{\Omega}$.

Substituting these relations into Eq.~\eqref{friction_nondissfinal_2D:eq}, we
obtain the 2D analogue of Eq.~\eqref{force_delta:eq}:
\begin{widetext}
\begin{align} \label{force_delta_2D:eq}
%\nonumber
  F(v_\text{SL})= &\frac{2\pi}{ m a^2} \sum_{l_y} (-1)^{l_y}
  \sum_{\lambda} \int_{\ell_\lambda^{\Omega}}
  \frac{dQ}{(2\pi)^2} \, Q_x \,
%  \times \\\nonumber&
  \tilde{V} (\left| \mathbf Q \right|) \, 
  \tilde{V} (\left| \mathbf Q + \mathbf G_\perp \right|)\,
  e^{-W(\mathbf{Q})-W(\mathbf{Q}+\mathbf{G}_\perp)}
%  \times \\ \label{force_delta_2D:eq} &
  \frac{ \mathbf Q \cdot \boldepsilon_{\lambda}(\mathbf Q) \,
    (\mathbf Q +\mathbf G_\perp) \cdot
    \boldepsilon_{\lambda}(\mathbf{Q})} 
    {\omega_{\lambda}(\mathbf{Q})\,| \mathbf{v}_\text{SL} -
      \mathbf{v}_\lambda(\mathbf{Q})|}
  \,.
\end{align}
\end{widetext}

The denominator $|\mathbf{v}_\text{SL} - \mathbf{v}_\lambda(\mathbf{Q})|$
vanishes at the special slider speeds, that, in addition to
Eq.~\eqref{resonance:eq}, also match the ``shock wave'' condition
\eqref{shock-wave:eq}.
In 2D this condition requires that both components of the group velocity $\mathbf{v}_{\lambda}(\mathbf{Q})$ match those of the slider velocity
\begin{align}\label{2D_shockwave_x:eq}
v_{\lambda\,x}(\mathbf{Q})\equiv\frac{\partial\omega_\lambda(\mathbf{Q})}{\partial
Q_x} &= v_\text{SL}\,, %\text{ and}
\\\label{2D_shockwave_y:eq}
v_{\lambda\,y}(\mathbf{Q})\equiv\frac{\partial\omega_\lambda(\mathbf{Q})}{\partial Q_y}&=0 \,.
\end{align}
The vanishing denominator $|\mathbf{v}_\text{SL} - \mathbf{v}_\lambda(\mathbf{Q})|$
gives the friction force an especially large singular
contribution.
Like the Van Hove singularities in the density of the electronic states and of
the harmonic phonon frequencies \cite{vanhove53}, in the 1D model a similar
singularity in the force integration led to peaks with diverging friction
\cite{Panizon18}; in the 2D model at hand these singularities generate
friction peaks characterized by a jump in the first derivative
$dF/dv_\text{SL}$; in 3D they lead to smoother and milder friction
peaks, as shown in Figs.~\ref{cube_center:fig} and \ref{bond_center:fig}.

\begin{figure} 
\centerline{
\includegraphics[width=0.8\linewidth,clip=]{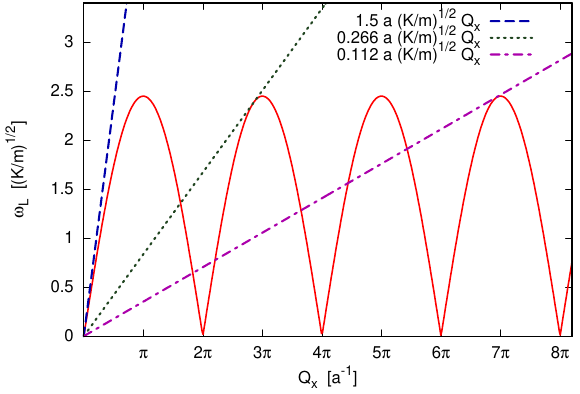}
}
\caption{\label{tangency_Qy0:fig}
Solid curve: the longitudinal phonon dispersion for the 2D simple-cubic crystal along the $Q_y=0$ line in a repeated-zone scheme.
Dashed line: the straight line whose slope is a supersonic speed
$v_\text{SL}=1.5 \, a(K/m)^{1/2} > v_\text{L}^\text{2D}$ never intersects the dispersion curve.
Other straight lines with slope $v_\text{SL}=0.266 \,
a(K/m)^{1/2}$ (dotted) and $v_\text{SL}=0.112 \, a(K/m)^{1/2}$ (dot-dashed)
are examples of speeds that meet the shockwave tangency
conditions \eqref{2D_shockwave_x:eq} and \eqref{2D_shockwave_y:eq}.
Note that a smaller $v_\text{SL}$ leads to a tangency point at a larger $Q_x$.
%
%{\bf NICK: this figure hints that when vSL is small we should worry about truncating the integration at a (not huge) $Q^\text{max}$,  because we miss very many shockwave contributions.  On the other hand, there the potential FT is supersmall, so these points don't really matter.}
%
}
\end{figure}

\begin{figure} 
\centerline{
\includegraphics[width=0.8\linewidth,clip=]{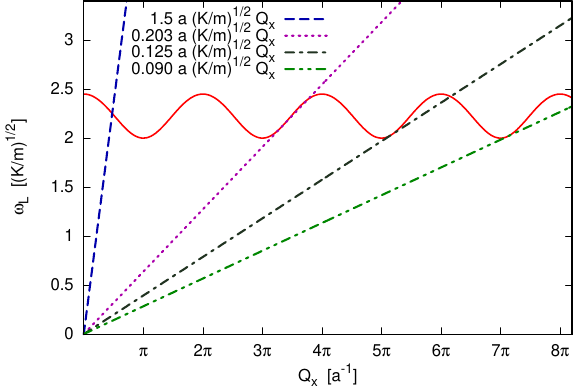}
}
\caption{\label{tangency_QyPi:fig}
Same as Fig.~\ref{tangency_Qy0:fig}, but along the $Q_y=\pi/a$ line.
Dashed line: the straight line at supersonic speed $v_\text{SL}=1.5 \, a(K/m)^{1/2}$ intersects the dispersion curve, but never meets the tangency condition \eqref{2D_shockwave_x:eq}.
Other straight lines with slope $v_\text{SL}
%=0.20257\, a(K/m)^{1/2}
\simeq 0.203\, 
a(K/m)^{1/2}$ (dotted), $v_\text{SL}\simeq 0.125\, 
a(K/m)^{1/2}$ (dot-dashed), and $v_\text{SL}\simeq 0.090 \, a(K/m)^{1/2}$ (dot-dot-dashed) provide examples of speeds for which the shockwave tangency
conditions are met.
}
\end{figure}

For the square-lattice phonon dispersions at hand, the condition
\eqref{2D_shockwave_y:eq} is verified only for $Q_y=n \pi/a$, for
integer $n$, indicating that shock-wave contributions, if any, can arise
only along these special symmetry lines.
As for the $Q_x$ component, locations where also condition
\eqref{2D_shockwave_x:eq} is fulfilled occur within suitable $v_\text{SL}$ ranges:
in particular, for small $Q_x$ and supersonic slider speed the shockwave condition \eqref{2D_shockwave_x:eq} is certainly {\em not} met.
Instead, this condition is met at larger and larger special $Q_x$ values, corresponding to smaller and smaller special $v_\text{SL}$ values.
Geometric examples of matching the shockwave condition \eqref{2D_shockwave_x:eq} are illustrated in Fig.~\ref{tangency_Qy0:fig} for even $n$, and in Fig.~\ref{tangency_QyPi:fig} for odd $n$.

As was done for the 3D model with Eq.~\eqref{friction_dissfinal:eq},
the introduction of a finite-phonon-lifetime $\gamma$ replaces the Dirac deltas with Lorentzian peaks, yielding
%\begin{equation}
%    \begin{split}
%        F(v_\text{SL})=& \frac{2}{ m a^2} \sum_{ l_y} (-1)^{l_y} \int_{\Omega'} \frac{d^2Q}{(2\pi)^2} \mathbf{Q} \cdot \hat{\mathbf{v}}_\text{SL} \tilde{V} (\left| \mathbf Q \right|) \tilde{V} (\left| \mathbf Q + \mathbf G_y \right|) e^{-W(\mathbf{Q})}e^{-W(\mathbf{Q}+\mathbf{G}_y)}  \\& \sum_{\lambda} \frac{1}{ \omega_{\lambda}(\mathbf{Q})} \mathbf Q \cdot \mathbf{\epsilon_{\lambda}(\mathbf Q)} \, (\mathbf Q +\mathbf G_y) \cdot \mathbf{\epsilon_{\lambda}(\mathbf{Q})} \frac{\frac{\gamma}{2}}{(\mathbf{Q}\cdot \mathbf{v}_\text{SL} - \omega_{\lambda}(\mathbf{Q}))^2-(\frac{\gamma}{2})^2} \, ,
%    \end{split}
%\end{equation}
%
\begin{widetext}
\begin{align} \label{friction_dissfinal_2D:eq}
  F(v_\text{SL})&=
  \frac{2\pi}{m a^2} \sum_{l_y} (-1)^{l_y}
  \int_\Omega \frac{d^2Q}{(2\pi)^2}
  \, Q_x \, 
%  \\\nonumber  &\times
  \tilde{V}(\left| \mathbf Q \right|) \, 
  \tilde{V} (\left| \mathbf Q + \mathbf G_\perp \right|) \, 
  e^{-W(\mathbf{Q})-W(\mathbf{Q}+\mathbf{G}_\perp)} \times 
  \\ \nonumber  
  & \hspace{42mm} \sum_{\lambda}
  \mathbf{Q} \cdot\boldepsilon_{\lambda}(\mathbf{Q}) \,
    (\mathbf{Q} +\mathbf{G}_\perp)\cdot
    \boldepsilon_{\lambda}(\mathbf{Q})\,
%  \\\nonumber  &\times
  \mathcal{L}(\mathbf{Q}, v_\text{SL}, \gamma)
  \,,
\end{align}
\end{widetext}
where $\mathcal{L}(\mathbf{Q}, v_\text{SL}, \gamma)$ is defined in
Eq.~\eqref{doubleLorenz_final:eq}.
%
%Where $\Omega'$ is given by the condition $(Q_x \in \mathbb{R},Q_y \geq 0)$.
Expression \eqref{friction_dissfinal_2D:eq} is the 2D analogue of Eq.~\eqref{friction_dissfinal:eq}, and is similarly ready to be evaluated numerically.

\subsection{Comparison with simulations}

Like in the 3D model we evaluate friction in the zero-temperature limit,
replacing the Debye-Waller factors with unity.
Like in Sect.~\ref{3D:sec}, $\hbar$ drops out of the expression, thus our theory predicts negligible quantum effect in this limit.

Even though the dispersion relation is available analytically, see Eq.~\eqref{omega_2D:eq}, in the code we compute $\omega_\lambda(\mathbf{Q})$ and the corresponding polarization vectors $\boldepsilon_{\lambda}$ by diagonalizing numerically the $2\times 2$ dynamical matrix defined by Eqs.~\eqref{2D_dynmatdiag:eq} and \eqref{2D_dynmatoffd:eq}, setting $K'=K/2$.
Like for the 3D model, in the figures below we report the friction force $F$ divided by the square of the perturbatively small coupling energy $\varepsilon$.

The code for the 2D model follows the same strategy as the 3D code, with the same real-space resolution $d_x=0.15\,a$ as in 3D.
For the integration grid we can afford a spacing $\Delta Q =
Q^\text{max}/N_Q$, with $N_Q=2500$, far finer than in 3D.
Thanks to this finer grid, we can adopt a smaller damping rate $\gamma =0.05\, (K/m)^{1/2}$, allowing us to investigate
the velocity dependence of friction in greater detail.
For the $\mathbf{G}_\perp$ summation, we consider $-15\leq l_y \leq 15$.
We verified that, thanks to the rapid decay of the Fourier transform of the potential and the smoothness of the integrand function, the evaluated friction force does not change visibly if these truncation limits and the grid fineness are improved.

\begin{figure} 
\centerline{
\includegraphics[width=\linewidth,clip=]{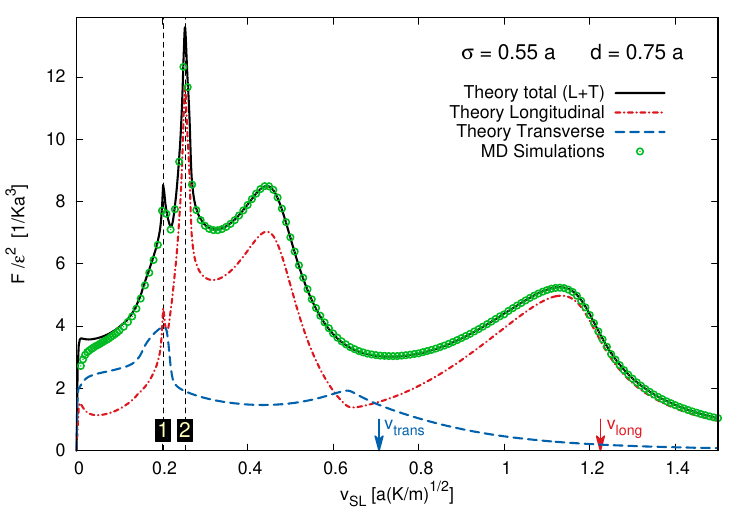}
}
\caption{\label{confronto_2D:fig}
Comparison of the slider-speed dependence of the friction force $F$
obtained evaluating the 2D LR expression \eqref{friction_dissfinal_2D:eq} (solid curve), with that obtained by
numerical MD simulations based on a crystal of $201 \times 201$ atoms, see
Appendix~\ref{MD:appendix} for details.
Dashed and dot-dashed curves: the contributions to the total friction of
the transverse and longitudinal phonons, respectively.
%
%Red and blue arrows: the crystal longitudinal and transverse speeds of sound, respectively.
%
Thin vertical dashed lines labeled 1 and 2: the speeds where friction peaks for reasons discussed in the text.
% and $v_\text{SL}= 0.203\, a(K/m)^{1/2}$
The reported friction force is divided by $\varepsilon^2$, expressing the
strength of the slider-crystal interaction potential, Eq.~\eqref{potr:eq}.
In both theory and simulations, a damping rate $\gamma = 0.05 \,
(K/m)^{1/2}$ is applied.
}
\end{figure}

\begin{figure} 
\centerline{
\includegraphics[width=\linewidth,clip=]{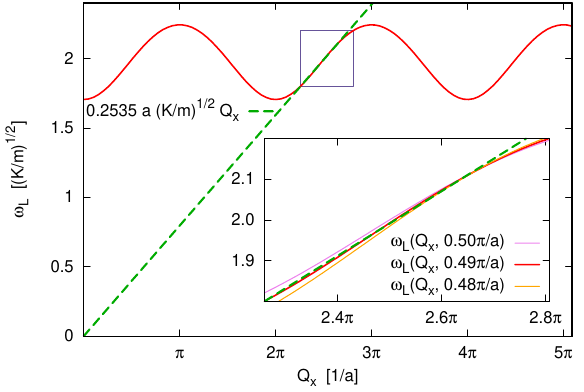}
}
\caption{\label{02535:fig}
Solid line: longitudinal phonon dispersion along the line
$Q_y\equiv 0.49 \frac{\pi}{a}$, compared to the straight line with slope
$v_\text{SL}=0.2545 \,a(K/m)^{1/2}$ (dashed).
Inset: a blowup of the close-approach region, illustrating how the straight
line associated to this specific speed comes extraordinarily close to the phonon dispersion over a sizable range of $Q_x$ and $Q_y$ values.
This approach is very close to satisfying
the resonance condition \eqref{resonance:eq}, but not the shockwave
condition, and specifically not Eq.~\eqref{2D_shockwave_y:eq}, which holds at $Q_y= n \frac{\pi}a$ only.
}
\end{figure}

\rem{useless:
\begin{figure} 
\centerline{
\includegraphics[width=\linewidth,clip=]{fig/v0203.pdf}
}
\caption{\label{0203:fig}
  Similar to Fig.~\ref{02535:fig}, but interpreting the weaker friction
  peak at $v_\text{SL}=0.203 \,a(K/m)^{1/2}$.
  Note that for this velocity, over a sizable range of $Q_x$, the phonon
  dispersion comes close to satisfying the resonance condition
  \eqref{resonance:eq}, and even the shockwave conditions
  \eqref{2D_shockwave_x:eq} \eqref{2D_shockwave_y:eq}, since $Q_y= \pi/a$.       
}
\end{figure}
}

Figure~\ref{confronto_2D:fig} compares the friction force obtained from the
evaluation of Eq.~\eqref{friction_dissfinal_2D:eq} (solid curve) with that
obtained by MD simulations (circles), executed with the
same parameters considered for the analytic expression, as detailed in
Appendix~\ref{MD:appendix}.
The qualititive and quantitative agreement is striking.
The analytic expression \eqref{friction_dissfinal_2D:eq} reproduces the MD
simulation result in all its detailed structure at all velocities, except
for small deviations at the lowest ones.
The red dot-dashed and blue dashed curves report the separate
contributions of the longitudinal and transverse phonons.
The longitudinal phonons dominate friction across the velocity range, except at relatively low velocity $v_\text{SL} < 0.2 \, a(K/m)^{1/2}$ %{\bf GABRIELE: 
and $0.6  \,a(K/m)^{1/2}<v_\text{SL}<0.7  \, a(K/m)^{1/2}$.

Friction exhibits a sharp peak,
labeled ``1'' in Fig.~\ref{confronto_2D:fig}, 
at $v_\text{SL}= 0.203 \, a(K/m)^{1/2}$.
As demonstrated by the dotted line in Fig.~\ref{tangency_QyPi:fig}, this is precisely the speed matching a tangency shockwave condition for the longitudinal phonon, but with substantial contributions from the transverse phonons, too. 

The next, even stronger, friction peak labeled ``2'' at
$v_\text{SL}= 0.2535\, a(K/m)^{1/2}$, marked by a vertical thin dashed line,  does {\em not} the result from a shockwave point instead, because at this velocity the tangency conditions \eqref{2D_shockwave_x:eq} and \eqref{2D_shockwave_y:eq}
are fulfilled nowhere in $\mathbf Q$ space, see Figs.~\ref{tangency_Qy0:fig} and \ref{tangency_QyPi:fig}.
We can rather understand this peak as generated by a relatively broad region of
$\mathbf Q$ satisfying approximately the resonance condition
\eqref{resonance:eq} for the longitudinal phonon dispersion near
% OLD, PROBABLY WRONG:
%$\mathbf Q \simeq (2.5,1)\,\frac{\pi}a$,
$\mathbf Q \simeq (2.6,0.49)\,\frac{\pi}a$,
see
Fig.~\ref{02535:fig}.
Although resonant matching is not exact, the adopted Lorentzian broadening makes the factor
$\mathcal{L}(\mathbf{Q}, v_\text{SL}, \gamma)$ in
Eq.~\eqref{friction_dissfinal_2D:eq} quite large over this wide $\mathbf Q$ region, contributing to make friction peak 2 so strong.

This same mechanism of an extended region of approximate resonance contributes to the intensity of peak 1, but instead lacks for the other shockwave resonances shown in Figs.~\ref{tangency_Qy0:fig} and \ref{tangency_QyPi:fig}.
This explains why these other resonances do not produce visible peaks for the considered phonon broadening.
If one could decrease $\gamma$, thus approaching the conservative-crystal limit $\gamma\to 0$, one would observe the appearance of multiple sharp (albeit relatively weak) shockwave peaks; at the same time the nonresonant peak 2 would progressively lose its prominence.

\begin{figure} 
\centerline{
\includegraphics[width=\linewidth,clip=]{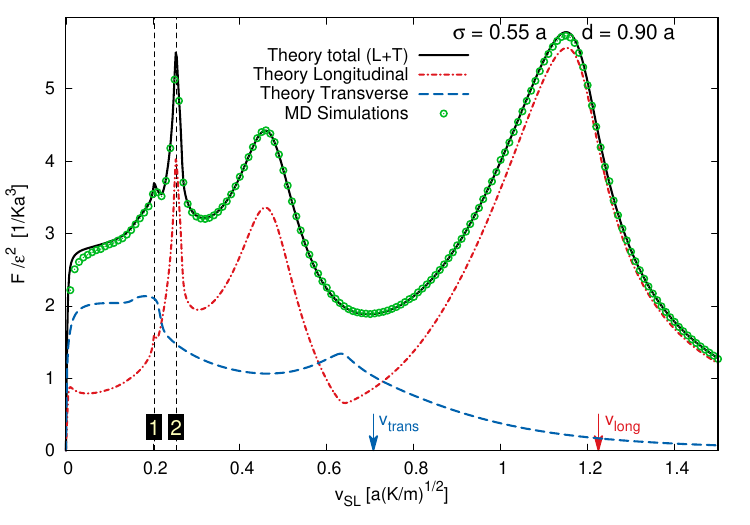}
}
\caption{\label{component_2D09:fig}
  Same as Fig.~\ref{confronto_2D:fig}, but for a different slider-crystal
  interaction, defined by $\sigma= 0.55 a$ and $d=0.9\, a$.
  Note that the friction peaks labeled 1 and 2 occur at the same speeds $v_\text{SL}= 0.203\, a(K/m)^{1/2}$ and $v_\text{SL}= 0.2535\,  a(K/m)^{1/2}$, as in Fig.~\ref{confronto_2D:fig}.
  This is expected, since the peaks positions only depend on the phonon dispersion, not on the slider-crystal interaction, which is the only different ingredient for the two graphs.
  In contrast, the relative fractional contributions to friction of  longitudinal and transverse phonons, as well as the absolute friction value, are significantly affected by changes in $V(r)$.
}
\end{figure}

Figure~\ref{component_2D09:fig} reports the friction force evaluated through
Eq.~\eqref{confronto_2D:fig}, but with a different slider-crystal interaction $V(r)$, defined by $\sigma= 0.55 \,a$ and $d=0.9\, a$, shown in Fig.~\ref{potx:fig}.
The accord between theory and simulations is as remarkable as for the other
interaction.
Observe that friction is weaker overall,
and that the speeds of friction peaks 1 and 2
are the same as in the calculation done with a different potential, reported in Fig.~\ref{confronto_2D:fig}.
This outcome is precisely predicted by our theory, which has friction peaks at velocities that depend uniquely on the phonon
dispersion, and not on the slider-crystal interaction.
Even though the peak positions are preserved, the detailed dependence of
friction on velocity is affected significantly by the interaction potential.

\section{Discussion and Conclusion} \label{discuss:sect}

Based on quantum linear-response (LR) theory complemented by well-understood approximations, we bring the theory of kinetic friction of a particle interacting with a periodic surrounding in the weak-interaction regime to an analytic formula, which is both formally instructive and
practically useful.
We derive concrete implementations of the general expression for particle channeling through two models of harmonic crystals: (i) a 3D simple-cubic crystal, and (ii) a 2D square lattice.

For both models, we provide a physical decomposition of the friction force
into a reciprocal-space integration and a summation of products of the
equilibrium dynamical response of the unperturbed crystal (a quantity
depending uniquely on the crystal phonon dispersions and polarizations and
on temperature) times the Fourier-transformed slider-crystal interaction.
This formulation shows that the advancing slider preferentially excites
the phonons which match the resonance condition, Eq.~\eqref{resonance:eq}:
these phonons are therefore identified as the main responsible for energy
dissipation.

We compare the 3D and 2D expressions with the friction evaluated by means
of MD simulations.
The remarkably close match serves as a validation of the analytical results.

Beside providing a conceptual understanding of the microscopic friction mechanism, the analytical formulation has obvious advantages over MD simulations to simulate friction.
For each sliding speed, the analytical friction calculation
requires the numerical evaluation of an integral, a task computationally far cheaper than setting up and running a full-fledged numerical simulation.
Additionally, the analytic result directly addresses the thermodynamic limit of a macroscopically large crystal, whereas simulations are confronted by size and time limitations, which could prove rather difficult to control, especially for 3D crystals.

%{\bf NICK: QUALCOS'ALTRO DA DIRE QUI? PER ESEMPIO HO PENSATO LE COSE SEGUENTI:
%In a real-life channeling experiment, sliders are {\em quantum} (as opposed to classical) particles.  
%Then perhaps it makes little or no sense imagining that they can follow a sharply defined path as in our model.
%So maybe it would make sense to average over infinitely many parallel paths, weighted with their specific probability, possibly related to the average interaction potential energy along that path.
%And potentially even this approach may be an undue simplification because these paths might even quantum interfere!
%And even non-straight paths should be considered and might also interfere...
%Mr. Feynman could have something to say here!
%In a completely different approach, the slow quantum motion perpendicular to channeling could be treated approximately independently of that in the channeling direction, as is done in Ref.~\cite{Korotchenko11}...}

The present results, though remarkable, are limited in scope by two main assumptions: (i) an infinite crystal, and (ii) a weak particle-crystal interaction.
The first assumption implies the rather idealized channeling geometry, as opposed to the more conventional contact-to-surface geometry, as, e.g., in ordinary AFM experiments.

As it stands, our theory for the energy loss in channeling could apply
e.g.\ to the slowing of neutral particles \footnote{Commonly channeled
charged particles \cite{LandauEMv8_ch14,Gemmell74,Logan92,Scandale21}
dissipate more energy through the excitation of electrons than through the
generation of phonons that is addressed by the present theory.}, e.g.\ fast
neutrons channeled across crystals
\cite{Kozhevnikova18,Baryshevskii90,Korotchenko11, Bee88}.
A reader who wished to apply the present theory to compute the energy loss
of channeled neutrons, though, should be warned that the extremely
short-range nature of the neutron-nucleus interaction would make the
relevant FT ${\tilde V}(q)$ decay extremely slowly.
In turn, this would make the numerical $\mathbf Q$ integration in Eq.~\eqref{friction_dissfinal:eq} more challenging, relying explicitly on the regularization provided by the large-$|\mathbf Q|$ decay of the Debye-Waller factors.

To address the very practical problem of evaluating analytically friction in a real-life sliding contact-to-surface or surface-surface interface, the next step will be an extension of this theory to contacts involving semi-infinite 3D crystals.
This extension can draw inspiration from Green's function approaches
\cite{Adelman76,BPBFV2005}.
For example, a Green's function approach has been successfully adapted to
take into account of the energy carried by classical phonons away from a
sliding contact and across semi-infinite bulk crystals \cite{Kajita10,
  Kajita16,Lee21,Tao24,Tao25},
and it has been recently combined with {\it ab-initio} simulations to take
the full chemistry of the sliding interface into account \cite{Kajita23}.
The current limitation of certain of those Green's function approaches lies
in their reliance on classical MD simulations for the atomic motions at the
interface, which lead to the well-known difficulty of fully exploring the
velocity scale, notably the low-speed region relevant for AFM experiments
\cite{Dong13,VanossiRMP13}.
An extension of the present linear-response analytic approach is expected to fully address this drawback.
For such an envisaged extension, one specific complication will be the need to describe surface phonons accurately, due to their likely central role for dissipation.
%
%It may be possible to address this problem by a suitable adaptation of techniques such as the one used in Ref.~\cite{Lee21}.
%
Further applications could require extension of this theory to general crystals and to non-crystalline solids, where not all simplifications based on symmetry used in the present work will be applicable.

The second main assumption, that of weak coupling, is appropriate for noncontact AFM experiments.
It will be important to overcome this limit when using this approach to address e.g.\ contact-mode AFM, in strong-friction conditions, where stick-slip dynamics arises.
Perhaps a Dyson-like resummation approach \cite{Kamenev11} could allow us
to extend this theory beyond LR.
Further aspects that call for future attention are temperature and quantum effects,
plus a realistic phonon anharmonicity to replace the phenomenological damping $\gamma$ in the theory.
These effects are nevertheless expected to be much more important for low-speed, stick-slip friction than in the present case.

\begin{acknowledgments}
We acknowledge useful discussion and feedback from Emanuele Panizon. 
E.T. and G.E.S. acknowledge that the research was partly supported by EU Horizon 2020 under ERC-ULTRADISS, Grant Agreement No. 834402.
G.E.S. acknowledges financial support from PNRR MUR project PE0000023-NQSTI, from PRIN 2022H77XB7 of the Italian Ministry of University and Research, and from the QuantERA II Programme STAQS project that 
has received funding from the European Union’s H2020 research and innovation programme under Grant Agreement No 101017733.
G.E.S. acknowledges that his research has been conducted within the framework of the Trieste Institute for Theoretical Quantum Technologies (TQT).

\end{acknowledgments}

\appendix

\section{The lattice phonons and their symmetries}
\label{lattice:app}

Our model elastic crystal consists of point particles with mass $m$ that at equilibrium are arranged as a square (2D) or simple-cubic (3D) lattice.
Harmonic nearest- and second-neighbor (diagonal) springs with, respectively, elastic constants $K$ and $K'$ and equilibrium lengths $a$ and $\sqrt{2} a$ guarantee the mechanical stability and determine the phonon properties.

The phonon polarization vectors $\boldepsilon_\lambda(\mathbf{q})$ and
frequencies $\omega_\lambda(\mathbf{q})$ are obtained by diagonalizing the
dynamical matrix $D_{\mu \nu}(\mathbf{q})$.
For these monoatomic crystals, the eigenvalue equation 
\begin{equation}\label{seculat:eq}
  \omega_\lambda^2(\mathbf{q}) \epsilon_{\lambda\,\mu}(\mathbf{q})
  = \sum_\nu D_{\mu \nu}(\mathbf{q}) \epsilon_{\lambda\,\nu}(\mathbf{q})
\end{equation}
involves a $2\times 2$ (2D) or $3\times 3$ (3D) matrix.

In 2D, the dynamical matrix has the following diagonal ($\mu\mu$) and
off-diagonal ($\mu\nu$) elements:
\begin{align}\nonumber
  D_{\mu\mu}(\mathbf{q}) =&
  \frac{2K}{m} \left[1-\cos{\left(q_\mu a\right)}\right]
  \\\label{2D_dynmatdiag:eq}
  &+\frac{2K'}{m}[1-\cos{(q_\mu a)}\cos{(q_\nu a)}]\, ,
  \\\label{2D_dynmatoffd:eq}
  D_{\mu\nu}(\mathbf{q}) =& \frac{2K'}{m}\sin{(q_\mu a)}\sin{(q_\nu a)} \,,
  \\\nonumber 
  \text{with }&\mu,\nu = x,y \text{ and }\nu \neq \mu\,.
\end{align}
From the secular equation \eqref{seculat:eq} we obtain the dispersion
relations:
\begin{align}\nonumber
  \omega_\pm(\mathbf{q}) =& \biggl\{
  2\frac{K}{m}\biggl[\sin^2{\left(\frac{q_xa}{2}\right)}
    +\sin^2{\left(\frac{q_ya}{2}\right)}
    \\ &
    +\frac{K'}{K}(1-\cos{(q_xa)}\cos{(q_ya})) \biggr]\nonumber
  \\\nonumber & \pm 2\frac{K}{m}\biggl[\left(
    \sin^2{\left(\frac{q_xa}{2}\right)}-\sin^2{\left(\frac{q_ya}{2}\right)}
    \right)^2
    \\\label{omega_2D:eq} &
    + \left(\frac{K'}{K}\sin{(q_xa)}\sin{(q_ya)}\right)^{\!2\,}
    \biggr]^{\nicefrac 12}
  \biggr\}^{\nicefrac 12}
  ,
\end{align}
for the $\lambda=1$ transverse ($-$) and $\lambda=2$ longitudinal ($+$)
phonons.
The (isotropic) transverse and longitudinal speeds of sound are 
\begin{align}\label{transv_speed_sound2D:eq}
  v^\text{2D}_\text{T} &=  a\left(\frac {K'}m \right)^{1/2}
  & \left[ = \frac1{\sqrt{2}}\, a\left(\frac Km \right)^{1/2} \right]
  ,
  \\\label{long_speed_sound2D:eq}
  v^\text{2D}_\text{L} &= a\left(\frac {K+K'}m \right)^{1/2} 
  & \left[ = \sqrt{\frac 32}\, a\left(\frac Km \right)^{1/2} \right]
  ,
\end{align}
with the expressions in square brackets corresponding to the adopted ratio
$K'/K=\frac 12$.

Likewise in 3D, the $3\times 3$ dynamical matrix can be expressed as an
explicit function of $\mathbf q$.
The diagonal elements are:
\begin{align}\label{dinamical_matrix_3Ddiag:eq}
  D_{\mu \mu}(\mathbf{q}) =& 2\frac{K}{m} [1-\cos{(q_\mu a)}]
  \\\nonumber
  &+ 2 \frac{K'}{m} [2- \cos(q_\mu a) (\cos(q_\nu a) +\cos(q_\gamma a))]
% CORRECTED FACTOR 2 ABOVE AND REMOVED WRONG FORMULA BELOW. AN ERROR WHICH MADE US SUFFER LONGTIME...
%\\\label{dinamical_matrix_3Doffd:eq}
%D_{\mu \nu}(\mathbf{q}) =&  4 \frac{K'}{m} \sin{(q_\mu a)} \sin{(q_\nu  a)}
%  \, ,
  \\\nonumber 
  \text{with }&\mu,\nu, \gamma = x,y,z \,,\ \nu \neq \mu\
  \text{ and }\gamma \neq \mu\,,\ \gamma \neq \nu \,.
\end{align}
The 2D expression for the off-diagonal elements \eqref{2D_dynmatoffd:eq} holds in 3D too.
Like in 2D, the diagonalization of $D$ provides normal-mode frequencies and
phonon polarization vectors for $\mathbf{q}$ spanning the first Brillouin
zone.
Diagonalization of the small-$|\mathbf{q}|$ expansion of this dynamical matrix determines the sound velocities.
At variance with the 2D case, they depend on the direction of $\mathbf q$.
For example in the $(100)$ direction, the transverse and longitudinal speeds of
sound are

\begin{align}\label{transv_speed_sound3D:eq}
  v^\text{3D}_\text{T} &=  a\left(\frac { K'}m \right)^{1/2}
  &\left[ = 
  \frac 1{\sqrt 2}\,
  a\left(\frac Km \right)^{1/2} \right]
  ,
  \\\label{long_speed_sound3D:eq}
  v^\text{3D}_\text{L} &= a\left(\frac {K+2 K'}m \right)^{1/2} 
  &\left[ = \sqrt{2}\, a\left(\frac Km \right)^{1/2} \right]
  ,
\end{align}
with the expressions in square brackets corresponding to the adopted value
$K'=K/2$.
For this same condition, in the (111) direction the sound velocities are
$$v^\text{3D}_\text{T}=\sqrt{\frac 23}a\left(\frac Km \right)^{1/2} \text{ and }
v^\text{3D}_\text{L}=\sqrt{\frac 53} a\left(\frac Km \right)^{1/2},$$ instead.

In both 2D and 3D, the dynamical matrix is explicitly even under inversion $\mathbf{q}\to -\mathbf{q}$.
As a result, $ \omega_\lambda(-\mathbf{q}) = \omega_{\lambda}(\mathbf{q})$.
We can also assume that the arbitrary phases of the polarization vectors
are fixed in such a way that
\begin{equation}\label{parity_polar:eq}
  \boldepsilon_\lambda(-\mathbf{q})
  = \boldepsilon_{\lambda}(\mathbf{q})
  \,.
\end{equation}

Additionally, we derive the symmetry of the polarization vectors under a
single-component reflection: $q_\mu \to -q_\mu$ (the two other components
$q_\nu$ and $q_\gamma$ remaining unchanged).
We indicate with $\mathbf{q}'$ the vector obtained from $\mathbf{q}$ under
this reflection.
By analyzing the secular equation \eqref{seculat:eq} at $\mathbf{q}'$, we
obtain the relevant symmetry relations for the polarization vectors:
\begin{align}\label{dynm_refl_symm}
  \omega_\lambda^2(\mathbf{q}') \epsilon_{\lambda\,\mu}(\mathbf{q}') =&
  D_{\mu \mu}(\mathbf{q}') \epsilon_{\lambda\,\mu}(\mathbf{q}')
  \\\nonumber
  &+ D_{\mu \nu}(\mathbf{q}') \epsilon_{\lambda\,\nu}(\mathbf{q}')
  +  D_{\mu    \gamma}(\mathbf{q}') \epsilon_{\lambda\,\gamma}(\mathbf{q}') 
  \\\nonumber
  =& D_{\mu \mu}(\mathbf{q}) \epsilon_{\lambda\,\mu}(\mathbf{q}')
  \\\nonumber
  &- D_{\mu \nu}(\mathbf{q}) \epsilon_{\lambda\,\nu}(\mathbf{q}')
  - D_{\mu \gamma}(\mathbf{q})\epsilon_{\lambda\,\gamma}(\mathbf{q}')
\\\nonumber
%  =& \omega_\lambda^2(\mathbf{q}) \epsilon_{\lambda\,\mu}(\mathbf{q}) \, .
  \text{with }& \gamma \neq \nu \neq \mu \neq \gamma \,.
\end{align}
Equation~\eqref{dynm_refl_symm} is satisfied provided that 
\begin{align}\label{omega_refl_symm}
  \omega_\lambda(\mathbf{q}') = \omega_{\lambda}(\mathbf{q})\,,
\end{align}
and that the corresponding polarization vector components are taken as
follow:
\begin{align}\label{epsilon_refl_symm}
  \epsilon_{\lambda\,\mu}(\mathbf{q}') &=
  \epsilon_{\lambda\,\mu}(\mathbf{q})
\\\nonumber
  \epsilon_{\lambda\,\nu}(\mathbf{q}') &=
  -\epsilon_{\lambda\,\nu}(\mathbf{q})\,,
  \text{ for }\nu\neq\mu\,.
\end{align}
This relation \eqref{epsilon_refl_symm} expresses the symmetry of the
polarization vectors under reflection in both 3D and 2D: in the calculation
of the lattice response function, Appendix~\ref{correlation:app}, we rely on this symmetry.

\section{Symmetries of the response function}\label{symm:sec}

Here we discuss a few symmetry property of the retarded response function
relevant for the calculation of friction.
Consider the Fourier representation of the retarded response function for
an arbitrary operator $A=A(\mathbf{r})$.
Clearly, if the two momentum arguments are the same, the real part is even
and the imaginary part is odd under $\mathbf{Q}\to-\mathbf{Q}$, namely:
\begin{align}
  \RE \chi_{AA}^R(\mathbf Q,\mathbf Q,\omega) &=
  \RE \chi_{AA}^R(-\mathbf Q,-\mathbf Q,\omega)
  \\
  \IM\chi_{AA}^R(\mathbf Q,\mathbf Q,\omega) &= -\IM
  \chi_{AA}^R(-\mathbf Q,-\mathbf Q,\omega)
\,.
\end{align}
However, for different momenta, as in Eq.~\eqref{power:eq}, no
straightforward parity symmetry holds.
However, for the specific periodic problem, where the arguments differ by a
$\mathbf G$ vector, another relevant symmetry property holds.

Consider the Fourier transform of a general correlation function for
operator $A$:
\begin{align}
  \chi_{AA}^R(\mathbf Q,\mathbf Q+\mathbf G,\omega)
  =& \int d^3\mathbf{x} \int d^3 \mathbf{x}' \times
  \\\nonumber &
  e^{-i\mathbf{Q} \cdot \mathbf{x}}
  \chi_{AA}^R(\mathbf x,\mathbf x',\omega) e^{i(\mathbf{Q} + \mathbf{G})
    \cdot \mathbf{x}'}
  \,.
\end{align}
Add to this expression the corresponding formula with $-\mathbf{G}$ in
place of  $\mathbf{G}$:
\begin{align}
  \chi_{AA}^R(\mathbf Q,&\mathbf Q+\mathbf G,\omega) + \chi_{AA}^R(\mathbf
  Q,\mathbf Q-\mathbf G,\omega) =
  \\\nonumber
  =&
  \int d^3\mathbf{x} \int d^3 \mathbf{x}' \,\times
  \\\nonumber&
  e^{-i\mathbf{Q} \cdot
    \mathbf{x}}\, \chi_{AA}^R(\mathbf x,\mathbf x',\omega) \left[
    e^{i(\mathbf{Q} + \mathbf{G}) \cdot \mathbf{x}'} +
    e^{i(\mathbf{Q} - \mathbf{G}) \cdot \mathbf{x}'} \right]
  .
\end{align}
If we replace $\mathbf{Q}$ with $-\mathbf{Q}$, the exponential factors at
the right hand side turn into their complex conjugate.
As a consequence,
\begin{align}
  \chi_{AA}^R&(\mathbf Q,\mathbf Q+\mathbf G,\omega) + \chi_{AA}^R(\mathbf
  Q,\mathbf Q-\mathbf G,\omega) =
  \\\nonumber
  &= \left[\chi_{AA}^R(-\mathbf Q,-\mathbf Q+\mathbf G,\omega) +
    \chi_{AA}^R(-\mathbf Q,-\mathbf Q-\mathbf G,\omega) \right]^{\dagger}
  .
\end{align}
We conclude that the real and imaginary parts of the $\mathbf
G$-symmetrized combination $\chi_{AA}^R(\mathbf Q,\mathbf Q+\mathbf
G,\omega) + \chi_{AA}^R(\mathbf Q,\mathbf Q-\mathbf G,\omega)$ are even and
odd functions of $\mathbf{Q}$, respectively.

%SPOSTATA DAL MAIN TEXT
We now make use of these symmetry properties for the evaluation of Eq.~\eqref{power:eq}.
We enucleate two subsets of the ${\mathbf G}_\perp$ vectors: $g^+$ including
the ${\mathbf G}_\perp$ vectors with both Miller indexes $l_y,l_z \ge 0$,
and $g^-$ including those with $l_y>0$ and $l_z <0$.
To include the remaining $l_y<0$ terms, we write:
\begin{align}
  \sum_{\mathbf G_\perp} &\chi_{nn}^R(\mathbf Q,\mathbf Q+\mathbf
  G_\perp,\mathbf Q \cdot \mathbf v_\text{SL}) =
  \\\nonumber
  =&  \sum_{\mathbf G_\perp\in g^+}
  \big[  \chi_{nn}^R(\mathbf Q,\mathbf Q+\mathbf G_\perp,\mathbf Q \cdot
    \mathbf v_\text{SL})
  \\\nonumber & \qquad
  +\chi_{nn}^R(\mathbf Q,\mathbf Q-\mathbf G_\perp,\mathbf Q \cdot \mathbf
  v_\text{SL}) \big] +
  \\\nonumber
  +&
  \sum_{\mathbf G_\perp\in g^-} \big[ \chi_{nn}^R(\mathbf Q,\mathbf Q+\mathbf
    G_\perp,\mathbf Q \cdot \mathbf v_\text{SL})
  \\\nonumber&\qquad
  + \chi_{nn}^R(\mathbf Q,\mathbf Q-\mathbf G_\perp,\mathbf Q \cdot \mathbf
  v_\text{SL})\big] \,.
\end{align}

This summation does not include all ${\mathbf G}_\perp$-dependent factors
in Eq.~\eqref{power:eq}: we must restore the initial-condition phase factor
and the Fourier transform of the interaction potential.
In fact, neither of these terms affects this symmetry property:
the phase factor is independent of $\mathbf Q$, therefore it cannot
change this symmetry;
$\tilde{V}(|\mathbf{Q}|)$ is a real function. We conclude that:
\begin{align}\label{Xisymmfin:eq}
  \chi_{nn}^R&(\mathbf Q,\mathbf Q+\mathbf G_\perp,\mathbf Q \cdot \mathbf
  v_\text{SL})\tilde{V} (\left| \mathbf Q + \mathbf G_\perp \right|)
  +
  \\ \nonumber
  \chi_{nn}^R&(\mathbf Q,\mathbf Q-\mathbf G_\perp,\mathbf Q \cdot \mathbf
  v_\text{SL}) \tilde{V} (\left| \mathbf Q - \mathbf G_\perp \right|) =
  \\ \nonumber &
  [\chi_{nn}^R(-\mathbf Q,-\mathbf Q+\mathbf G_\perp,-\mathbf Q \cdot
    \mathbf v_\text{SL})\tilde{V} (\left| - \mathbf Q + \mathbf G_\perp
    \right|)
    + \\ \nonumber &
    \ 
    \chi_{nn}^R(-\mathbf Q,-\mathbf Q-\mathbf G_\perp,-\mathbf Q \cdot
    \mathbf v_\text{SL})\tilde{V} (\left| -\mathbf Q - \mathbf G_\perp
    \right|)]^*
%{\dagger} ????
  \,.
\end{align}

\section{Retarded linear-response function}\label{correlation:app}

Here we derive expressions for the density-density response function, first
for the standard conservative harmonic crystal, then for a crystal
characterized by dissipative phonons.
%Its importance is evident because it is the most important part of the
%retarded response function, Eq.~\eqref{chi:eq}.?????? hahhaha!

\subsection{Conservative crystal}\label{non_diss:subsubsec}

We start from a regular conservative crystal.
We express the retarded LR function
%Eq.~\eqref{chi_def:eq},  NOTE THAT Riva references a different equation!!!
\begin{equation}\label{chi:eq}
  \chi_{nn}^R(\mathbf x,\mathbf x',t-t') \equiv
  -\frac{i}{\hbar}\theta(t-t')\langle [ n(\mathbf x,t-t'),
    n(\mathbf x',0)]\rangle
\end{equation}
in its Fourier representation:
% NICK: In the following, for brevity,
% I replace "\omega" in place of "\mathbf Q \cdot \mathbf v_\text{SL}"
\begin{align}\label{chi_FT:eq}
  \chi_{nn}^R&(\mathbf Q,\mathbf Q+\mathbf G_\perp,\omega)
  \\\nonumber &
  = \lim_{V \to \infty} -\frac{i}{\hbar} \int dt\, \theta(t) e^{i\omega t}
  \frac{1}{V}\int_V d^3x \int_V d^3x'
  \\\nonumber &
  \quad\times
  e^{-i\mathbf Q \cdot \mathbf{x}} e^{i(\mathbf Q + \mathbf{G_\perp}) \cdot
    \mathbf{x}'} \left \langle  \left[ n(\mathbf x,t), n (\mathbf
    x',0) \right] \right\rangle
  \\\nonumber &
  = \lim_{V \to \infty} - \frac{i}{\hbar V} \int dt\, \theta (t) e^{i\omega t}
  \\\nonumber &
  \quad\times
  \left \langle  \left[ n(\mathbf Q,t), n (-\mathbf Q -\mathbf G_\perp,0)
    \right] \right\rangle
  \,.
\end{align}
%{\bf NICK: vedo un'asimmetria tra il primo e il secondo argomento della
%  chi. Questo introduce un segno opposto nei wavevectors.  Perche`?}
Here we introduced
\begin{equation}
  n(\mathbf Q,t)=
  \int_V d^3x \, e^{-i\mathbf Q \cdot \mathbf{x}} \, n(\mathbf x,t) \, ,
\end{equation}
the spatial Fourier transform of the density operator.
As customary for the vibrations of a crystal, we express the position
operator of the $j$-th atom at time $t$ as
\begin{equation}  % perhaps inline
  \mathbf{x}_j=\mathbf{R}_j+\mathbf{u}_j(t) \,,
\end{equation}
the sum of its equilibrium position $\mathbf{R}_j$ plus its displacement
operator $\mathbf{u}_j(t)$.
We substitute this decomposition into the definition \eqref{densop:eq} of
the density operator, obtaining
\begin{align}\label{nQt:eq}
  n(\mathbf Q,t)&=
  \int_V d^3x \, e^{-i\mathbf Q \cdot \mathbf{x}} \,\sum_{j=1}^N
  \delta^{(3)}(\mathbf{x} - \mathbf{x}_j)
  \\\nonumber &
  = \sum_{j=1}^N e^{-i\mathbf Q \cdot \mathbf{R}_j}e^{-i\mathbf Q \cdot
    \mathbf{u}_j(t)}
  \,.
\end{align}
$N$ is the number of atoms in the crystal, hence the volume $V=Na^3$.
Periodic boundary conditions are assumed to preserve crystalline symmetry.
In 2D, the Dirac deltas and the $\mathbf x$-integration of Eq.~\eqref{nQt:eq}
become two-dimensional.

Equation~\eqref{chi_FT:eq} involves the density-density correlation
function in the Fourier domain:
\begin{widetext}
\begin{align}\nonumber
  \langle  n (\mathbf Q,t)\, n (-\mathbf Q -\mathbf G_\perp,0)\rangle =
&
%\\\nonumber &  = 
\sum_{j,j'}
  e^{-i  \mathbf Q \cdot (\mathbf R_j - \mathbf R_{j'}) }
  \left \langle e^{-i \mathbf u_j(t) \cdot \mathbf Q } e^{i \mathbf
    u_{j'}(0) \cdot (\mathbf Q +\mathbf G_\perp) } \right \rangle 
\\ = &
   N \sum_j  e^{-i  \mathbf Q \cdot \mathbf R_j } \left \langle e^{-i
    \mathbf u_j(t) \cdot \mathbf Q } e^{i \mathbf{u}_0(0) \cdot (\mathbf Q
    +\mathbf G_\perp) } \right \rangle
  ,
\end{align}
where $\mathbf{u}_0$ is the displacement operator of an atom at an
arbitrarily fixed site, say the origin  $\mathbf{R}_0 = \mathbf 0$.
To evaluate these quantum averages, we use the Gaussian identity
\begin{equation}\label{gaussianHO}
  \langle e^A e^B   \rangle =
  e^{\frac{1}{2}\langle  A^2 \rangle + \frac{1}{2}\langle  B^2  \rangle
    +\langle  A \, B \rangle}
\end{equation}
valid for harmonic-oscillator operators \cite{Mermin66}.
The direct application of Eq.~\eqref{gaussianHO} to our case of interest,
leads to:
\begin{align}\label{division:eq}
\left\langle e^{-i \mathbf{u}_j(t) \cdot \mathbf Q }
e^{i \mathbf u_0(0) \cdot (\mathbf Q +\mathbf G_\perp) } \right\rangle
=
e^{-\frac{1}{2} \langle \mathbf u_j(t) \cdot \mathbf Q \,
  \mathbf u_j(t) \cdot \mathbf Q  \rangle }
  \,
e^{-\frac{1}{2} \langle \mathbf{u}_0(0) \cdot
  (\mathbf{Q}+\mathbf{G}_\perp)\, \mathbf{u}_0(0) \cdot (\mathbf{Q}
    +\mathbf{G}_\perp) \rangle }
\,e^{ \langle \mathbf{u}_j(t) \cdot \mathbf Q \, \mathbf{u}_0(0)
  \cdot (\mathbf Q + \mathbf{G}_\perp) \rangle }
\,.
\end{align}
\end{widetext}
To evaluate these averages, we express the displacement operators in terms
of phonon annihilation $b_{\mathbf{k}\lambda}$ and creation
$b_{\mathbf{k}\lambda}^{\dagger}$ operators \cite{Ashcroft}
\begin{align}\label{displacement_eq}
  \mathbf{u}_j(t)
  =& \frac{1}{\sqrt{N}}
  \sum_{\mathbf{k},\lambda} \sqrt{\frac{\hbar}{2 m
      \omega_{\lambda}(\mathbf{k})}}
  \boldepsilon_{\lambda}(\mathbf{k})
  e^{i \mathbf{k} \cdot \mathbf{R}_j}
  \\\nonumber &\times
  \left( e^{-i\omega_{\lambda}(\mathbf{
      k})t}\, b_{\mathbf{k}\lambda} + e^{i\omega_{\lambda}(\mathbf{k})t}
  \, b_{-\mathbf{k}\lambda}^{\dagger}  \right)
    .
\end{align}
We start calculating the equal-time quantum average 
\begin{align}\nonumber
  \langle  &\mathbf{u}_j(t) \cdot \mathbf Q \, \mathbf{u}_j(t) \cdot
  \mathbf Q  \rangle
  = \langle \mathbf{u}_0(0) \cdot \mathbf Q
  \, \mathbf{u}_0(0) \cdot \mathbf Q  \rangle 
  \\ \nonumber 
  &= \frac{1}{N} \sum_{\mathbf{k},\mathbf{k'}}\sum_{\lambda,\lambda'}  \mathbf{Q} \cdot \boldepsilon_{\lambda}(\mathbf{k}) \mathbf{Q} \cdot \boldepsilon_{\lambda}(\mathbf{k'}) 
  \frac{\hbar}{2m\sqrt{\omega_{\lambda}(\mathbf{k})\omega_{\lambda'}(\mathbf{k'})}} \nonumber \\
  &\times \Big \langle \left(e^{-i \omega_{\lambda'}(\mathbf{k'})t}
  b_{\mathbf k' \lambda'} + e^{i \omega_{\lambda'}(\mathbf{k'})t}
  b^{\dagger}_{-\mathbf k' \lambda'} \right)\nonumber \\
  &\times \left(e^{-i \omega_{\lambda}(\mathbf{k})t}b_{\mathbf k \lambda} +  e^{i \omega_{\lambda}(\mathbf{k})t}b^{\dagger}_{-\mathbf k \lambda} \right)  \Big \rangle = \nonumber \\
  &= \frac{1}{N} \sum_{\mathbf{k},\mathbf{k'}}\sum_{\lambda,\lambda'}  \mathbf{Q} \cdot \boldepsilon_{\lambda}(\mathbf{k}) \mathbf{Q} \cdot \boldepsilon_{\lambda}(\mathbf{k'}) 
  \frac{\hbar}{2m\sqrt{\omega_{\lambda}(\mathbf{k})\omega_{\lambda'}(\mathbf{k'})}} \nonumber \\
  &\times    \Big [ e^{-i (\omega_{\lambda}(\mathbf{k})-\omega_{\lambda'}(\mathbf{k'}))t} \left \langle
  b^{\dagger}_{-\mathbf k' \lambda'} b_{\mathbf k \lambda}\right
  \rangle \nonumber \\
  &+ e^{i (\omega_{\lambda}(\mathbf{k})-\omega_{\lambda'}(\mathbf{k'}))t} \left \langle 
  b_{\mathbf k' \lambda'} b^{\dagger}_{-\mathbf k \lambda} \right
  \rangle  \Big],
  \nonumber 
\end{align}
calculating the quantum average between creation and annihilation operators as 
\begin{align}\label{quantum_everage:eq}
    \left \langle
  b^{\dagger}_{-\mathbf k' \lambda'} b_{\mathbf k \lambda}\right
  \rangle &=\delta_{\mathbf{k}, -\mathbf{k'}}\delta_{\lambda,\lambda'}
  n_{\lambda}(\mathbf{k}) 
  %\eta_{\lambda}(\mathbf{k}) 
  \\
  \left \langle 
  b_{\mathbf k' \lambda'} b^{\dagger}_{-\mathbf k \lambda} \right
  \rangle &= \delta_{\mathbf{k}, -\mathbf{k'}}\delta_{\lambda,\lambda'}
  (1+n_{\lambda}(\mathbf{k}))\nonumber 
  %(1+\eta_{\lambda}(\mathbf{k}))\nonumber 
\end{align}
and using the parity in $\mathbf{k}$ of the phonon frequency
\eqref{parity_polar:eq} and of the polarization vectors we obtain
\begin{align}
  \langle  &\mathbf{u}_j(t) \cdot \mathbf Q \, \mathbf{u}_j(t) \cdot
  \mathbf Q  \rangle
  = \langle \mathbf{u}_0(0) \cdot \mathbf Q
  \, \mathbf{u}_0(0) \cdot \mathbf Q  \rangle \nonumber
  \\ \nonumber 
  &= \frac{1}{N} \sum_{\mathbf{k},\lambda} | \mathbf{Q} \cdot
  \boldepsilon_{\lambda}(\mathbf{k}) |^2
  \frac{\hbar}{2m\omega_{\lambda}(\mathbf{k})}(2n_{\lambda}(\mathbf{k})+1)
  \\ \label{debyewaller:eq}&
  \equiv 2W(\mathbf{Q})
\,,
\end{align}
and, likewise,
\begin{align}\label{other_debyewaller:eq} 
  \langle \mathbf{u}_0(0) & \cdot (\mathbf Q+\mathbf{G}_\perp)\,
  \mathbf{u}_0(0) \cdot (\mathbf Q +\mathbf{G}_\perp) \rangle
  \equiv 2W(\mathbf{Q}+\mathbf{G}_\perp)
  \,.
\end{align}
Here $n_{\lambda}(\mathbf{k}) = 1/(e^{\beta \hbar
  \omega_{\lambda}(\mathbf{k}) }-1)$ is the average number of quanta in the
oscillator labeled by $\mathbf{k}$ and $\lambda$ at equilibrium at
temperature $T=1/(k_\text{B} \beta)$.
Since all Bose excitation factors $n_{\lambda}(\mathbf{k})$ increase with
temperature, so does the thermal vibration amplitude of the crystal, with
the result that the value of these Debye-Waller \cite{Ashcroft} factors
$e^{-W(\mathbf{Q})}$ decreases.
When the temperature becomes so high that $\beta \hbar
\omega_\lambda(\mathbf{k}) \ll 1$ the crystal is packed with phonons,
i.e.\ $n_\lambda(\mathbf{k}) \simeq k_\text{B} T /(\hbar
\omega_\lambda(\mathbf{k})) \gg 1 $, also $W(\mathbf{Q})$ is very large,
and, as a result, the Debye-Waller factor decays rapidly toward zero.
As a result, the friction force is also reduced: this ubiquitous effect,
named thermolubricity, was studied in several different contexts
\cite{Sang01,Dudko02,Szlufarska08,Brukman08,Steiner09}. 
Note that the Debye-Waller factors are similar for the two terms
generated by the commutator in Eq.~\eqref{chi_FT:eq}.

The third factor in Eq.~\eqref{division:eq} is the only one that explicitly
depends on time.
Therefore both elements of the commutator in Eq.~\eqref{chi_FT:eq} need to
be calculated.
We start evaluating the first one
\begin{align}
  \langle &  \mathbf{u}_j(t) \cdot \mathbf Q \,
  \mathbf{u}_0(0) \cdot (\mathbf Q + \mathbf G_\perp) \rangle
  \\\nonumber&
  = \frac{\hbar}{2mN}
  \sum_{\mathbf{k},\mathbf{k'}} \sum_{\lambda,\lambda'}   \frac{\mathbf Q \cdot
    \boldepsilon_{\lambda}(\mathbf{k})(\mathbf Q +\mathbf G_\perp)
    \cdot
    \boldepsilon_{\lambda'}(\mathbf{k'})}{\sqrt{\omega_{\lambda}(\mathbf
      k)\omega_{\lambda'}(\mathbf{k'})}} e^{i \mathbf{k'}\cdot \mathbf R_j
  }
  \\\nonumber&
  \times  \left \langle \left(e^{-i \omega_{\lambda'}(\mathbf{k'})t}
  b_{\mathbf k' \lambda'} + e^{i \omega_{\lambda'}(\mathbf{k'})t}
  b^{\dagger}_{-\mathbf k' \lambda'} \right)\left(b_{\mathbf k \lambda} +
  b^{\dagger}_{-\mathbf k \lambda} \right) \right \rangle
  \\\nonumber &
  = \frac{\hbar}{2mN} \sum_{\mathbf{k},\mathbf{k'}} \sum_{\lambda,\lambda'}
  \frac{\mathbf Q \cdot \boldepsilon_{\lambda}(\mathbf k)(\mathbf Q
    +\mathbf G_\perp) \cdot
    \boldepsilon_{\lambda'}(\mathbf{k'})}{\sqrt{\omega_{\lambda}(\mathbf
      k)\omega_{\lambda'}(\mathbf{k'})}} e^{i \mathbf{k'}\cdot \mathbf R_j
  }
  \\\nonumber &
  \times   \left [ e^{i \omega_{\lambda'}(\mathbf{k'})t} \left \langle
  b^{\dagger}_{-\mathbf k' \lambda'} b_{\mathbf k \lambda}\right
  \rangle  + e^{-i \omega_{\lambda'}(\mathbf{k'})t} \left \langle 
  b_{\mathbf k' \lambda'} b^{\dagger}_{-\mathbf k \lambda} \right
  \rangle  \right].
  \nonumber 
\end{align}
By evaluating the quantum average as in Eq.~\eqref{quantum_everage:eq} and
using the parity in $\mathbf{k}$ of the phonon frequency and of the
polarization vector, as analyzed in Appendix~\ref{lattice:app}, we obtain
\begin{align}\label{corr_time:eq}
  \langle &  \mathbf{u}_j(t) \cdot \mathbf Q \,
  \mathbf{u}_0(0) \cdot (\mathbf Q + \mathbf G_\perp) \rangle
  \\\nonumber
  &= \frac{\hbar}{2mN}
  \sum_{\mathbf{k}} \sum_{\lambda}   \frac{\mathbf Q \cdot
    \boldepsilon_{\lambda}(\mathbf k)(\mathbf Q +\mathbf G_\perp)
    \cdot
    \boldepsilon_{\lambda}(\mathbf{k})}{\omega_{\lambda}(\mathbf k)}
  \\\nonumber &
  \times e^{-i \mathbf{k} \cdot \mathbf R_j }
  \left [ e^{i
    \omega_{\lambda}(\mathbf{k})t} n_{\lambda}(\mathbf k) + e^{-i
      \omega_{\lambda}(\mathbf{k})t}(1+n_{\lambda}(\mathbf k))
    \right]
  \\\nonumber &
  = \frac{\hbar}{2mN} \sum_{\mathbf{k}} \sum_{\lambda}
  \frac{\mathbf Q \cdot
    \boldepsilon_{\lambda}(\mathbf{k})\,
    (\mathbf Q +\mathbf G_\perp)\cdot\boldepsilon_{\lambda}(\mathbf{k})}
       {\omega_{\lambda}(\mathbf{k})}  e^{-i \mathbf{k} \cdot \mathbf R_j } 
  \\\nonumber& \times
  \left [(2n_{\lambda}(\mathbf
    k)+1) \cos(\omega_{\lambda}(\mathbf{k})t) - i
    \sin(\omega_{\lambda}(\mathbf{k})t) \right].
\end{align}
Note that, for a crystal lattice with different symmetries, not all
simplifications carried out in Eq.~\eqref{corr_time:eq} may apply.

Consider the other term $\left \langle \mathbf{u}_0(0) \cdot (\mathbf Q +
\mathbf G_\perp) \, \mathbf{u}_j(t) \cdot \mathbf Q \right \rangle$,
arising in the calculation of the
%correlator???
commutator in Eq.~\eqref{chi_FT:eq}, processed through
Eq.~\eqref{division:eq}.
With a similar derivation, we obtain an expression which is the complex
conjugate of the result of Eq.~\eqref{corr_time:eq}:
\begin{align}\label{corr_cc:eq}
  \langle& \mathbf{u}_0(0) \cdot (\mathbf Q + \mathbf G_\perp) \,
  \mathbf{u}_j(t) \cdot \mathbf Q \rangle
  \\\nonumber&
  = \frac{\hbar}{2mN} \sum_{\mathbf{k}} \sum_{\lambda}
  \frac{\mathbf Q \cdot
    \boldepsilon_{\lambda}(\mathbf{k})\,
    (\mathbf Q +\mathbf G_\perp)\cdot\boldepsilon_{\lambda}(\mathbf{k})}
       {\omega_{\lambda}(\mathbf{k})}  e^{-i \mathbf{k} \cdot \mathbf R_j }
  \\\nonumber& \times
  \left [(2n_{\lambda}(\mathbf{k})+1) \cos(\omega_{\lambda}(\mathbf{k})t)
    + i \sin(\omega_{\lambda}(\mathbf{k})t) \right]
  .
\end{align}

At this point, it is convenient to define
\begin{align}\label{phi_def:eq}
  \phi_j(t,\beta)&\equiv
  \left \langle \mathbf{u}_j(t) \cdot \mathbf Q \,
  \mathbf{u}_0(0) \cdot (\mathbf Q + \mathbf G_\perp) \right \rangle
  \\\nonumber&
  = \left \langle \mathbf{u}_0(0) \cdot (\mathbf Q + \mathbf G_\perp)\,
  \mathbf{u}_j(t) \cdot \mathbf Q \right \rangle^\dagger
  .
\end{align}
Observe that the real part of $\phi_j(t,\beta)$ depends both on time and on
temperature, while its imaginary part is independent of temperature:
\begin{align}\label{im_re_phi:eq}
  \phi_j(t,\beta)&= \RE \phi_j(t,\beta) + i \IM \phi_j(t,\beta)
    \\\nonumber&
    = \phi^\text{r}_j(t,\beta)+i \phi^\text{i}_j(t)
    \,.
\end{align}
This same observation applies also for the 1D counterpart of this formula
\cite{Panizon18}.

We have now all elements required to write an explicit expression for the
Fourier representation of the relevant density-density response function,
Eq.~\eqref{chi_FT:eq}:
\begin{align}\label{chi_nonapprx:eq}
  \chi_{nn}^R&(\mathbf Q,\mathbf Q+\mathbf G_\perp,\omega) =
  -\frac{i}{\hbar a^3}e^{-W(\mathbf{Q})} e^{-W(\mathbf{Q} +\mathbf{G}_\perp )}
  \\\nonumber  & \times
  \int dt \,\theta (t)\, e^{i \omega t}  \sum_{j=0}^{N-1}
  e^{-i\mathbf{Q}\cdot \mathbf{R}_j} 
%  \\\nonumber  & \times
  \left[ e^{\phi_j(t,\beta)}-  e^{\phi_j^{\dagger}(t,\beta)}\right]
  \\\nonumber &=
  -\frac{i}{\hbar a^3}e^{-W(\mathbf{Q})} e^{-W(\mathbf{Q} +\mathbf{G}_\perp )}
  \\\nonumber  & \times
  \int dt \,\theta (t)\, e^{i \omega t}  \sum_{j=0}^{N-1}
  e^{-i\mathbf{Q}\cdot \mathbf{R}_j} e^{ \phi^r_j(t,\beta)}
  \\\nonumber  & \times
  \left[ e^{i
     \phi^i_j(t)}-  e^{-i \phi^i_j(t)}\right]
  \\\nonumber &=
  \frac{2}{\hbar a^3} e^{-W(\mathbf{Q})} e^{-W(\mathbf{Q}
    +\mathbf{G}_\perp )}   \int dt \, \theta(t) \, e^{i\omega t}
  \\\nonumber  & \times
  \sum_{j=0}^{N-1}
  e^{-i\mathbf{Q}\cdot \mathbf{R}_j} e^{\phi^\text{r}_j(t,\beta)}
  \sin{(\phi^\text{i}_j(t))}
  \,.
\end{align}
This is an exact expression for the retarded density-density
LR function of the vibrating harmonic lattice.
Evaluating $ \chi_{nn}^R(\mathbf Q,\mathbf Q+\mathbf G_\perp, \omega)$
through Eq.~\eqref{chi_nonapprx:eq} requires carrying out a double
integration, namely over the $\mathbf k$ points in the BZ (in
$\phi_j(t,\beta)$) and over time, plus a summation over the lattice
translations, which makes it quite cumbersome and inefficient to compute
numerically.
Unfortunately, we cannot see a way to further simplify the final expression
Eq.~\eqref{chi_nonapprx:eq}, while keeping it exact.

As was done for the 1D problem \cite{Panizon18}, we resort to the
one-phonon approximation \cite{Ashcroft} for the terms involving
$\phi_j(t,\beta)$:
\begin{equation} \label{1phon:eq}
  e^{\phi^\text{r}_j(t,\beta)} \sin(\phi^\text{i}_j(t))
  \simeq \phi^\text{i}_j(t)
  \, ,
\end{equation}
which is valid in the limit of small $|{\phi_j(t,\beta)}|$, appropriate for
not too large wave vector $Q$ and not too high temperature.
%Sec.~\ref{1Dchi:sec}.
%
By inserting the one-phonon approximation \eqref{1phon:eq} in
Eq.~\eqref{chi_nonapprx:eq}, we obtain
\begin{align}\label{chiQ_pass1}
  \chi&_{nn}^{R\text{ 1 ph}} (\mathbf Q,\mathbf Q+\mathbf G_\perp,\omega)
  \\\nonumber &
  \simeq  % qui metterei il simeq, cosi` e` chiaro che da qui e` approssimato
  \frac{2}{\hbar a^3} \, e^{-W(\mathbf{Q})}
  e^{-W(\mathbf{Q} +\mathbf{G}_\perp )}
  \int dt \, \theta(t) \, e^{i\omega t}
  \sum_{j=0}^{N-1} e^{-i\mathbf{Q}\cdot \mathbf{R}_j}
  \\\nonumber &\times
  \frac{(-\hbar)}{2 m N}
  \sum_{\mathbf{k},\lambda}
  \frac{
  \mathbf Q \cdot
  \boldepsilon_{\lambda}(\mathbf k)\,
  (\mathbf Q +\mathbf G_\perp)\cdot
  \boldepsilon_{\lambda}(\mathbf{k})}{\omega_{\lambda}(\mathbf k)}
  \\\nonumber &\times
   e^{-i \mathbf{k} \cdot \mathbf R_j }
  \sin(\omega_{\lambda}(\mathbf{k})t)
  \\\nonumber &
  = % faccio un paio di semplificazioni importanti!
  -\frac{1}{m a^3 N} \, e^{-W(\mathbf{Q})}
  e^{-W(\mathbf{Q} +\mathbf{G}_\perp )}
  \int dt \, \theta(t) \, e^{i\omega t}
  \\\nonumber &\times
  \sum_{j=0}^{N-1} e^{-i\mathbf{Q}\cdot \mathbf{R}_j}
  \sum_{\mathbf{k},\lambda}
  \frac{
  \mathbf Q \cdot
  \boldepsilon_{\lambda}(\mathbf k)\,
  (\mathbf Q +\mathbf G_\perp)\cdot
  \boldepsilon_{\lambda}(\mathbf{k})}{\omega_{\lambda}(\mathbf k)}
  \\\nonumber &\times e^{-i \mathbf{k} \cdot \mathbf R_j }
  \sin(\omega_{\lambda}(\mathbf{k})t)
  \,.
\end{align}
Here, the simplification of Planck's constant indicates that in the
one-phonon approximation the only quantum effects occur through the
Debye-Waller factors.
Executing the thermodynamic limit we obtain:
\begin{align}\label{chiQ_pass2}
  \chi&_{nn}^{R\text{ 1 ph}}(\mathbf Q,\mathbf Q+\mathbf G_\perp,\omega)
  \\\nonumber &
  = -\frac{1}{m} e^{-W(\mathbf{Q})} e^{-W(\mathbf{Q}
    +\mathbf{G}_\perp )} \int dt\, \theta(t)\, e^{i \omega t}
  \sum_{j} e^{-i\mathbf{Q}\cdot \mathbf{R}_j}
  \\\nonumber &\times
  \sum_{\lambda} \int_{BZ} \frac{d^3k}{(2\pi)^3}\,
  \frac{\mathbf{Q} \cdot \boldepsilon_{\lambda}(\mathbf k)
  \,(\mathbf Q +\mathbf G_\perp)\cdot
  \boldepsilon_{\lambda}(\mathbf{k})}
  {\omega_{\lambda}(\mathbf k)}
  \\\nonumber &\times
    e^{-i \mathbf{k} \cdot \mathbf R_j }  \sin(\omega_{\lambda}(\mathbf{k})t) \, ,
\end{align}
where the integral over $\mathbf k$ extends over the first Brillouin zone, and the sum
over $j$ spans the infinitely-many translations $\mathbf{R}_j$ of
the Bravais lattice.
To simplify this expression we can use the identity
$\sin(x)=(e^{ix}-e^{-ix})/(2i)$
%{\bf NICK: this one is clearly applied to the sin term with time, all right}
%and $\int_{-a}^a dx \cos{(x)} f(x) =
%\int_{-a}^a dx e^{ix} f(x) $, 
%{\bf NICK: this one suggests application to the cos term, but it is not a 1D integral, there is no integration relative to the variable of the cos,and $a$ is an already used symbol...}
%valid if $f(x)$ is an even function.
%{\bf The fact that the rest of the integrated function (all except for
%cos) is even depends on the phonon polarization vectors being even
%functions, which we have not yet discussed at this point. Should we move that
%observation to yet another appendix????}
%
Accordingly,
\begin{align}\label{chiQ_pass3}
  \chi&_{nn}^{R\text{ 1 ph}}(\mathbf Q,\mathbf Q+\mathbf G_\perp,\omega)
  \\\nonumber &
  = \frac{i}{2m} e^{-W(\mathbf{Q})} e^{-W(\mathbf{Q}
    +\mathbf{G}_\perp )} \int dt \theta (t)
  \\\nonumber &\times
  \sum_{j} \sum_{\lambda} \int_{BZ} \frac{d^3k}{(2\pi)^3}  e^{-i \mathbf{R}_j \cdot (\mathbf{Q}-\mathbf{k})}
  \\\nonumber & \times
   \frac{\mathbf Q \cdot \boldepsilon_{\lambda}(\mathbf k)\,(\mathbf Q
     +\mathbf G_\perp) \cdot \boldepsilon_{\lambda}(\mathbf{k})}
        {\omega_{\lambda}(\mathbf k)}
  \\\nonumber & \times
   \left[e^{ it(\omega_{\lambda}(\mathbf{k}) + \omega)} -
    e^{-it(\omega_{\lambda}(\mathbf{k})- \omega)}\right]
  .
\end{align}
Now we make use of the 3D version of the periodic delta function, or
\textit{Dirac comb}, identity:
\begin{equation}\label{deltacomb}
  \sum_{j} e^{-i \mathbf{R}_j \cdot (\mathbf{Q}-\mathbf{k})} =
  \frac{(2\pi)^3}{a^3} \sum_{\mathbf{G}'}
  \delta^3(\mathbf{k} -\mathbf{Q} - \mathbf{G}') \, ,
\end{equation}
where $\mathbf{G}' =\frac{2\pi}{a} (l_x,l_y,l_z) $ are the reciprocal
lattice vectors of the Bravais lattice.
Hence, combining the integral over $\mathbf{k}$ extended over the first BZ
with the $\mathbf{G}'$ summation, we obtain an unrestricted integral on a
variable $\mathbf{k}'=\mathbf{k} - \mathbf{G}'$ spanning the the entire
reciprocal space.
Furthermore, the $\delta^3(\mathbf{k}'-\mathbf{Q})$ condition eliminates the integration over $\mathbf{k}'$, and fixes $\mathbf{k}'$ to $\mathbf{Q}$.
We rewrite Eq.~\eqref{chiQ_pass3} as
\begin{align}\label{time_int:eq}
  \chi&_{nn}^{R\text{ 1 ph}}(\mathbf Q,\mathbf Q+\mathbf G_\perp,\omega)
  \\\nonumber &
  = \frac{i}{2m a^3} e^{-W(\mathbf{Q})} e^{-W(\mathbf{Q}
    +\mathbf{G}_\perp )} 
  \\\nonumber &\times
  \sum_{\lambda}
  \frac{\mathbf{Q} \cdot \boldepsilon_{\lambda}(\mathbf
    Q)\,(\mathbf Q +\mathbf G_\perp) \cdot
  \boldepsilon_{\lambda}(\mathbf{Q})}{\omega_{\lambda}(\mathbf{Q})}
  \\\nonumber &\times
  \int dt\, \theta (t)
  \left[e^{it(\omega+\omega_{\lambda}(\mathbf{Q}))}
  -e^{it(\omega-\omega_{\lambda}(\mathbf{Q}))}\right]
  \,.
\end{align}
Since all factors in Eq.~\eqref{time_int:eq} except for the initial
imaginary unit and the final time integral are real, and we are interested
in the imaginary part of $\chi_{nn}^R$ only, we need to evaluate the real
part of the time integral only.
We recall the real part of the Fourier transform of the Heaviside $\theta$
function:
\begin{equation}
  \RE \int dt \, \theta (t) \, e^{i\omega t} = \pi \delta(\omega)
  \,.
\end{equation}
Accordingly,
\begin{align}\label{chi_immnond:eq}
  \IM\chi&_{nn}^{R\text{ 1 ph}}(\mathbf Q,\mathbf Q+\mathbf G_\perp,\omega) 
  \\\nonumber &=
  \frac{\pi}{2m a^3} e^{-W(\mathbf{Q})} e^{-W(\mathbf{Q}
    +\mathbf{G}_\perp )}
  \\\nonumber &\times
  \sum_{\lambda} \frac{\mathbf{Q} \cdot \boldepsilon_{\lambda}(\mathbf Q)
  \,(\mathbf Q +\mathbf G_\perp) \cdot
  \boldepsilon_{\lambda}(\mathbf{Q})
  }{\omega_{\lambda}(\mathbf{Q})}
  \\\nonumber &\times
  \left[\delta(\omega+\omega_{\lambda}(\mathbf{Q}) ) -
   \delta(\omega-\omega_{\lambda}(\mathbf{Q}) )\right]
  .
\end{align}
In Sect.~\ref{conservative:sec}, we adopt this approximate expression for
the calculation of friction.

\subsection{Dissipative crystal}\label{diss:subsubsec}

The next step is the introduction of a finite phonon lifetime.
For this purpose, we add a small imaginary part to the frequency: $\omega
\to \omega+i\gamma/2$.
This modification introduces a uniform exponential decay with rate $\gamma/2$ to all phonon modes, corresponding to a lifetime $2/\gamma$.
With this phenomenological modification, we rewrite the retarded
one-phonon-approximate LR function, Eq.~\eqref{time_int:eq},
as follows:
\begin{align}\label{chi_diss_time:eq}
  \chi&_{nn}^{R\text{ 1 ph}}(\mathbf Q,\mathbf Q+\mathbf G_\perp,\omega)
  \\\nonumber &
  = \frac{i}{2m a^3} \,
  e^{-W(\mathbf{Q})} e^{-W(\mathbf{Q} +\mathbf{G}_\perp)}
  \\\nonumber & \times
  \sum_{\lambda}
  \frac{\mathbf{Q} \cdot \boldepsilon_{\lambda}(\mathbf Q)(\mathbf Q
    +\mathbf G_\perp) \cdot
    \boldepsilon_{\lambda}(\mathbf{Q})}{\omega_{\lambda}(\mathbf Q)}
  \\\nonumber & \times
  \int dt \, \theta(t)
  (e^{it(\omega +\omega_{\lambda}(\mathbf{Q}) + i\frac{\gamma}{2})}
  -e^{it(\omega -\omega_{\lambda}(\mathbf{Q}) + i\frac{\gamma}{2})})
  \,.
\end{align}
%Compared to the infinite-lifetime phonons, where we rely on the Fourier
%transform of the Heaviside theta function, now the
The time integration is of this elementary form:
\begin{equation}
  \int_{-\infty}^{+\infty} dt \, \theta(t)\, e^{it(\omega + i\frac{\gamma}{2})}
  = \int_{0}^{+\infty} \, dt \,  e^{it(\omega + i\frac{\gamma}{2} )}
  = \frac{i}{\omega + i\frac{\gamma}{2}}
  \,.
\end{equation}
Therefore, the time integral in the last line of
Eq.~\eqref{chi_diss_time:eq} becomes:
\begin{align}\nonumber
  &\int_{-\infty}^{+\infty} dt \theta(t)
  \left(e^{it(\omega+\omega_{\lambda}(\mathbf{Q}) + i\frac{\gamma}{2} )}
  -e^{it(\omega -\omega_{\lambda}(\mathbf{Q}) + i\frac{\gamma}{2} )}
  \right) 
  \\\label{chi_diss_time1} &=
  i \left( \frac{1}{\omega+\omega_{\lambda}(\mathbf{Q}) + i\frac{\gamma}{2}}
  - \frac{1}{\omega-\omega_{\lambda}(\mathbf{Q}) + i\frac{\gamma}{2}}
  \right)
  \\\nonumber &=
  i \left( \frac{\omega+ \omega_{\lambda}(\mathbf{Q}) - i\frac{\gamma}{2}}
  {(\omega+ \omega_{\lambda}(\mathbf{Q}))^2 + (\frac{\gamma}{2})^2}
  - \frac{\omega-\omega_{\lambda}(\mathbf{Q})  - i\frac{\gamma}{2}}
  {(\omega -\omega_{\lambda}(\mathbf{Q}))^2 + (\frac{\gamma}{2})^2} \right)
  \\\nonumber &=
  \left( \frac{i(\omega+ \omega_{\lambda}(\mathbf{Q})) +\frac{\gamma}{2}}
  {(\omega+ \omega_{\lambda}(\mathbf{Q}))^2 + (\frac{\gamma}{2})^2}
  - \frac{i(\omega-\omega_{\lambda}(\mathbf{Q}))  +\frac{\gamma}{2}}
  {(\omega -\omega_{\lambda}(\mathbf{Q}))^2 + (\frac{\gamma}{2})^2} \right)
  .
\end{align}
In the last expressions we have separated the real and imaginary parts of
both fractions.
%multiplied and divided by $\pm \omega_{\lambda}(\mathbf{Q}) + \omega -
%i\frac{\gamma}{2}$ with $+$ and $-$ for the first and the second factor in
%brackets respectively.
All other factors in Eq.~\eqref{chi_diss_time:eq} being real except for the
initial imaginary unit, the imaginary part of $\chi_{nn}^{R\text{ 1 ph}}$
is obtained by retaining the real part of the result of the time
integration in Eq.~\eqref{chi_diss_time1}:
\begin{align}\label{chi_immd:eq}
  \IM \chi&_{nn}^{R\text{ 1 ph}}(\mathbf{Q},\mathbf{Q}+
  \mathbf{G}_\perp,\omega)
  \\\nonumber &=
  \frac{\pi}{2m a^3} e^{-W(\mathbf{Q})} e^{-W(\mathbf{Q} +\mathbf{G}_\perp)}
  \\\nonumber &\times  
  \sum_{\lambda}
  \frac{\mathbf Q \cdot \boldepsilon_{\lambda}(\mathbf Q)\,
  (\mathbf Q +\mathbf G_\perp) \cdot
  \boldepsilon_{\lambda}(\mathbf{Q})
  }{\omega_{\lambda}(\mathbf Q)}
  \\\nonumber &\times  
  \left[ \frac{\frac{\gamma}{2\pi}}{(\omega + \omega_{\lambda}(\mathbf{Q}))^2
    + (\frac{\gamma}{2})^2} -
  \frac{\frac{\gamma}{2\pi}}{(\omega -\omega_{\lambda}(\mathbf{Q}))^2
    + (\frac{\gamma}{2})^2} \right]
  .
\end{align}
A comparison of Eq.~\eqref{chi_immd:eq} with the infinite-lifetime
expression \eqref{chi_immnond:eq} shows that the introduction of
dissipation has the unique effect of broadening the Dirac
$\delta(\omega -\omega_{\lambda}(\mathbf{Q}))$ to a Lorentzian:
\begin{equation}\label{deltatogauss}
  \delta(\omega_{\lambda}(\mathbf{Q}) - \omega) \to
  \frac{\frac{\gamma}{2\pi}}{(\omega - \omega_{\lambda}(\mathbf{Q}))^2 
    +  (\frac{\gamma}{2})^2}
  \,.
\end{equation}
In Sect.~\ref{dissipative:sec}, we adopt the one-phonon expression
\eqref{chi_immd:eq} for the evaluation of friction in the dissipative
crystal.

\section{Molecular-dynamics simulations}\label{MD:appendix}

We simulate a cubic portion of the harmonic simple-cubic lattice consisting of $N= 51\times 51\times 51$ atoms.
For the 2D model, we consider a square portion of the harmonic square lattice involving $N
= 201\times 201$ atoms.
In both models, atoms are connected by nearest-neighbor springs (with strength $K$ and rest length $a$), and second-neighbor springs (with strength $K'=K/2$ and rest length $\sqrt{2} a$).
We apply periodic boundary conditions to mitigate finite-size effects.

We integrate $T=0$ Langevin equations of motion, with a viscous term with damping coefficient $\gamma$, adopting the values discussed in Sects.~\ref{3D:sec} and \ref{2D:sec}.
Numerical integration relies on an accurate Runge-Kutta-Feldberg algorithm that adapts its time step to maintain a predetermined accuracy.
The slider is kept advancing at the fixed speed $v_\text{SL}$ along a line halfway
between the rest positions of two adjacent atomic rows or through the center of a sequence of aligned cubes in the $(100)$ direction.
The slider and the crystal interact through the potential of
Eq.~\eqref{potr:eq},  with the weak coupling amplitude
$\varepsilon = 5\times 10^{-4}\,Ka^2$.
In the simulations, we adopt a smoothly truncated version of $V(r)$, which
vanishes beyond a cutoff distance $5 a$.
We typically run simulations over a total time as large as $2\times 10^4\,
(m/K)^{1/2}$ for the lowest simulated speed $v_\text{SL} = 0.01 \, a
(K/m)^{1/2}$, then progressively reduced to $1\times 10^3 \, (m/K)^{1/2}$ for
$v_\text{SL}\geq 0.2 \, a (K/m)^{1/2}$, in such a way that the slider advances by at least 200 lattice spacings in each
simulation.

As the crystal starts off in its static equilibrium configuration, it needs
some time to adjust itself to the presence of the advancing slider.
We consider the simulation steps covering the initial $150 \, (m/K)^{1/2}$
time interval as a transient, and therefore we omit them from the
evaluation of steady-state average properties.
By monitoring the average kinetic energy of the crystal lattice, we
verified that the crystal average temperature never exceeds a tiny $T =
4\times 10^{-10}\, Ka^2/k_\text{B}$, indicating that simulations are
indeeed representative of the $T=0$ regime.

To compute the average friction force, in principle one could evaluate the time average of the force experienced by the slider.
However, this procedure is numerically quite challenging, since the slider force in the advancing direction oscillates widely, with instantaneous values of the order of $\varepsilon/\sigma$.
This fluctuation is several orders of magnitude larger than its seeked-for average value which, according to Eq.~\eqref{frictionOoM}, is very small, of order $\varepsilon^2$.
Therefore we rather evaluate $F$ by means of a numerically more stable and
convenient procedure: The advancing slider pushes the whole crystal forward
by transferring to it a rightward momentum $\mathbf{F}(t) dt$.
As a result, in the initial transient the center of mass of the crystal
accelerates, but it soon acquires a nearly steady velocity
$\mathbf{V}_\text{CM}$ when the viscous force $-m \gamma \mathbf{v}_j$ that
the Langevin thermostat applies to each crystal atom generates a total
compensating leftward momentum transfer $\mathbf{F}_\text{diss}(t) dt =
-m \gamma \sum_j \mathbf{v}_j dt$.
This compensation is not exact at each instant of time: as a result, the
crystal center of mass advances to the right at a not-exactly-constant
velocity.
However, in the steady state the fluctuations of $\mathbf{V}_\text{CM} =
N^{-1} \sum_j \mathbf{v}_j $ are a tiny fraction of its average value, due
to the huge inertia of the crystal ($N\gg 1$).
Moreover, for the adopted small coupling $\varepsilon$, the value of
$|\mathbf{V}_\text{CM}|$ is quite small, never exceeding $1.6\times
10^{-9} \, a (K/m)^{1/2}$.
In practice, the time-average of $\mathbf{V}_\text{CM}$ is obtained by the
numerically extremely stable and reliable expression
\begin{align}
  \overline{\mathbf{V}}_\text{CM} \equiv
  \frac 1{t_2-t_1} \int_{t_1}^{t_2} \mathbf{V}_\text{CM}(t) dt
  = \frac{\mathbf{X}_\text{CM}(t_2)-\mathbf{X}_\text{CM}(t_1)}{t_2-t_1}
  \,,
\end{align}
evaluated with $t_1$ at the end of the initial transient and $t_2$ at the
end of the simulation.
Given $\overline{\mathbf{V}}_\text{CM}$ from a simulation, then the average
friction force is evaluated as
\begin{align}
  \mathbf{F} = N m \gamma \overline{\mathbf{V}}_\text{CM}
  \,.
\end{align}
Of course, in all these vectors, the transverse components vanish: only the component in the $\hat{\mathbf{e}}_x$ sliding direction is
relevant.
For example, $F=F_x$ is the quantity reported as circles in
Figs.~\ref{cube_center:fig}, \ref{bond_center:fig}, \ref{confronto_2D:fig}, and \ref{component_2D09:fig}.

\section{3D Fourier Transform of the potential}\label{FT3d:appendix}
In 3D, the FT is defined and evaluated as follows:
\begin{align}\nonumber
 \tilde{V}(\mathbf{q})
  &=
  \int d^3x \, e^{-i\mathbf{q} \cdot \mathbf{x}} \, V(|\mathbf{x}|)
  \\\nonumber
  &=
  2\pi \int_0^{+\infty} dr r^2 V(r)\int_{-1}^{+1}d\cos{\theta}\,
  e^{-iqr\cos{\theta}} =
  \\\nonumber
  &= \frac{4\pi}{q} \int_0^{+\infty} dr\, r \,
  V(r)\sin{(qr)}
  \\\label{3DFTofV:eq}
  &=  \tilde{V}(|\mathbf{q}|) =  \tilde{V}(q)
  \,.
\end{align}
As noted in the final line, this sort of ``atomic form factor of the
slider'' is a real function, and it is spherically symmetric in reciprocal
space, i.e.\ $\tilde{V}$ only depends on the length of the $\mathbf{q}$
vector.

For evaluating the integral in Eq.~\eqref{3DFTofV:eq}, we first proceed to
manipulate it
using the relation $\sin(qr)=\frac{e^{iqr}-e^{-iqr}}{2 i}$.
Observe that the potential is even in $r$, which allows us to rewrite the
integral as
\begin{align}
    \int_{0}^{+\infty}dr\, r\, V(r)\frac{e^{iqr}-e^{-iqr}}{2 i}
    =\int_{-\infty}^{+\infty}dr\, r \, V(r)\frac{e^{iqr}}{2 i}.
\end{align}
For our regularized LJ function, Eq.~\eqref{potr:eq}: 
\begin{align}
    \tilde V(\mathbf{q})=&\frac{4 \pi\varepsilon}{q}\!\int_{-\infty}^{+\infty}\!
    dr\, r 
\\\nonumber    &\left[\left(\frac{\sigma^2+d^2}{r^2+d^2}\right)^{\!6}-2\left(\frac{\sigma^2+d^2}{r^2+d^2}\right)^{\!3}\right]\frac{e^{iqr}}{2 i}.
\end{align}
To evaluate this integral, we consider the two terms in the sum separately.
We start solving the simpler one
\begin{equation}
    \int_{-\infty}^{+\infty}dr \,r \left(\frac{1}{r^2+d^2}\right)^3e^{iqr}
\end{equation}
where all the constants are omitted.
This function has triple poles at $r=\pm i d$.
Notice that due to the presence of $e^{iqr}$ we close the integration in the upper half of the complex $r$ plane. 
We recall the general expression for a residue of order $n$ at location $c$:
\begin{equation}
    \text{Res}(f(z),c)=\frac{1}{(n-1)!}\lim_{z\to c}\,\frac{d^{n-1}}{dz^{n-1}}[(z-c)^nf(z)]
    \,.
\end{equation}
For our function and order 3:
\begin{align}
    \text{Res}&(f(z),id)=\frac{1}{2}\lim_{z\to id}\frac{d^{2}}{dz^{2}}\nonumber \\
    &\left[(z-id)^3 z\left(\frac{1}{(z+id)(z-id)}\right)^3e^{iqz}\right]\nonumber \\
    &=\frac{1}{2}\lim_{z\to id}\,\frac{d^{2}}{dz^{2}}(z\frac{1}{(z+id)^3}e^{iqz})\nonumber \\
    &=\frac{q}{16d^3}\,e^{-qd}(1+qd)
\end{align}
using the residue theorem, this part gives the term with $(qd+1)$ in Eq.~\eqref{pot_fourier:eq}.
With the same approach we calculate the remaining part, involving the integral
\begin{equation}
    \int_{-\infty}^{+\infty}dr\, r\, \left(\frac{1}{r^2+d^2}\right)^6e^{iqr}
    .
\end{equation}
The 6$^\text{th}$ order residue of this integrand is
%(mathematica did the derivative)
\begin{align}
    \text{Res}&(f(z),id)=\frac{1}{5!}\lim_{z\to id}\,\frac{d^{5}}{dz^{5}}(z\frac{1}{(z+id)^6}e^{iqz})\nonumber\\
    &=\frac{q\,e^{-qd}}{5! 2^6 d^9}
    [105+105qd+45q^2d^2+10q^3d^3+q^4d^4]\,,
\end{align}
that generates the remaining terms in Eq.~\eqref{pot_fourier:eq}.

\section{2D Fourier Transform of the potential}\label{FT2d:appendix}
The 2D Fourier transform of the interaction potential of
Eq.~\eqref{potr:eq} is:
\begin{align}\nonumber
\tilde{V}(q)&=\int d^2x \,e^{-i\mathbf{q}\cdot \mathbf{x}}\, V(|
\mathbf{x} |)
\\\nonumber
&= \int_0^{+\infty} dr \,r\, V(r) \int_0^{2\pi} d\theta \,
e^{-iqr\cos{\theta}}
\\\label{2D_pot_step:eq}
&= 2\pi \int_0^{+\infty} dr \, r J_0(qr)\,V(r)
\,,
\end{align}
where $J_0(\cdot)$ is the $0^\text{th}$-order Bessel function of the first kind.
%:\begin{equation}
%    J_0(x)= \frac{1}{2\pi}\int_0^{2\pi} d\theta e^{-ix\cos{\theta}} \,.
%\end{equation}

To derive an explicit expression for the integral in
Eq.~\eqref{2D_pot_step:eq}, we resort to Equation 10.22.46 of
Ref.~\cite{NIST:DLMF:integrals}, that we specialize to $\nu=0$ and
reformulate as follows:
\begin{equation}
    \int_0^{\infty}
    dr\,r\, J_0(qr)
    \frac{b^{2(n+1)}}{(r^2+b^2)^{n+1}}
    =
    \frac{b^2}{n!}
    \left(\frac{qb}2\right)^n
    K_n(qb)
    \,,
\end{equation}
where $K_n(\cdot)$ is a modified Bessel function of the second kind.
By applying this formula twice, namely with $n=5$ and $n=2$, we obtain the expression reported as Eq.~\eqref{2D_pot:eq}.

Note that the divergences of
the modified Bessel function of the second
kind $K_n(qd)$ for $qd\to 0$ are compensated by the corresponding power terms, so that the $qd\to 0$ limit of Eq.~\eqref{2D_pot:eq} is finite, namely:
\begin{align}\label{2D_potq0:eq}
    \tilde{V}(0)
    =
    2\pi \varepsilon d^2 
    \left[
    \frac 1{10}
    \left(1+\frac {\sigma^2}{d^2}\right)^6
    -\frac 12
    \left(1+\frac {\sigma^2}{d^2}\right)^3
    \right]
    .
\end{align}

In practice, the straightforward implementation of Eq.~\eqref{2D_pot:eq} based on the python {\tt scipy.special} standard package is numerically stable down to 
$qd\geq 2\times 10^{-9}$.
Below this value of $qd$, our implementation returns the result of Eq.~\eqref{2D_potq0:eq}.
%

%\bibliography{biblio}

\end{document}